\documentclass[%
 aip,pop,
 amsmath,amssymb,
 preprint,%
]{revtex4-1}

\usepackage{graphicx}% Include figure files
\usepackage{dcolumn}% Align table columns on decimal point
\usepackage{bm}% bold math
\usepackage{color}
\usepackage{hyperref}
\hypersetup{
     colorlinks   = true,
     citecolor    = blue,
     linkcolor     = blue}

\renewcommand{\v}[1]{\ensuremath{\mathbf{#1}}} % for vectors

\begin{document}

%\preprint{AIP/123-QED}

\title{Generalized slab universal instability and its appearance in pair plasma}

\author{Ben Zhu}%
 \email{zhu12@llnl.gov}
 \affiliation{Lawrence Livermore National Laboratory\\7000 East Avenue\\Livermore, California 94550-9234, USA}
\author{Manaure Francisquez}%
 \affiliation{Plasma Physics and Fusion Center, Massachusetts Institute of Technology, Cambridge, MA 02139 USA}
\author{Barrett N. Rogers}
 \affiliation{Department of Physics and Astronomy, Dartmouth College\\6127 Wilder Laboratory\\Hanover, New Hampshire 03755-3528, USA}
\author{Xue-qiao Xu}
 \affiliation{Lawrence Livermore National Laboratory\\7000 East Avenue\\Livermore, California 94550-9234, USA}

\date{\today}% It is always \today, today,

\begin{abstract}
A generalized linear dispersion relation of electromagnetic slab universal modes is derived, taking into account arbitrary ion charge state, electron finite Larmor radius (FLR) effects, and Debye shielding effects. As a consequence, it provides more accurate predictions and is applicable to a wider range of plasmas. We find that electron FLR effects have a weakly stabilizing effect on the slab universal instability in electron-ion plasma, while Debye shielding strongly stabilizes this instability when $\lambda_D$ approaches $\rho_i$ ($\lambda_D$ is the Debye length and $\rho_i$ is the ion gyro-radius). In particular, we examine the stability criterion for this instability in electron-positron pair plasmas, and find that the instability persists in this simplest plasma system as long as the pair plasma number density exceeds the critical value $n_c=B^2/(8\pi m_e c^2)$. 

%Valid PACS numbers may be entered using the \verb+\pacs{#1}+ command.
\end{abstract}

\maketitle

\section{Introduction}\label{sec:intro}
This work describes a linear gyrokinetic analysis of the electromagnetic universal instability in a collisionless slab plasma.
The mode we term a ``universal instability" in this research differs slightly from the the most restrictive usage of that term, and might equally well be labeled an entropy mode. Historically, the names ``entropy mode"~\cite{kadomtsev1960convective} and ``universal instability"~\cite{galeev1963universal} were first introduced in 1960s.
The former refers to an ion Larmor scale thermal instability driven by density and temperature gradients, resulting a perturbed density and temperature while the plasma kinetic pressure remains undisturbed. In certain magnetic configurations (e.g., Z-pinch~\cite{ricci2006gyrokinetic} and dipole~\cite{simakov2001kinetic}), it is believed to play an essential role in particle and heat transport when ideal MHD instabilities are suppressed.
For instance, it is found that the entropy mode is responsible for the observed particle pinch (i.e., transport of particles along the direction of the density gradient) in the dipole configuration with both local gyrokinetic~\cite{kobayashi2009gyrokinetic} and global fluid~\cite{ou2020turbulent} simulations.
The universal mode (instability) was first named to refer to the electrostatic instability predicted in low $\beta$ magnetized plasma occurring due to non-uniform densities -- a pervasive characteristic amongst almost all magnetized plasmas, and hence the term \textit{universal}. This name was challenged later as subsequent work focusing on the long wavelength limit ($k_\perp\rho_i\ll 1$) found that this instability is not truly ``universal" as it can be stabilized in complex geometries\cite{krall1965universal} or in the simple sheared slab in the absence of additional instability drivers (i.e., temperature gradients, parallel current or magnetic curvatures).~\cite{antonsen1978stability,ross1978drift,tsang1978absolute}
However, recent studies suggest that the universal instability could still exist at sub-ion Larmor scales $k_\perp\rho_i \geq 1$ in a sheared slab~\cite{smolyakov2002short,landreman2015universal} or in more general geometries~\cite{helander2015universal}.
To date, most research into entropy modes and the universal instability has been carried out in the electrostatic limit.

Previously, we performed a local linear electromagnetic gyrokinetic analysis of a shearless, collisionless slab plasma with the constraint of MHD equilibrium, i.e., equilibrium pressure balance $p+B^2/(8\pi)=\text{constant}$, and discovered an instability driven by density and temperature gradients.~\cite{rogers2018gyrokinetic} Allowing electromagnetic fluctuations while enforcing MHD equilibrium implies that the plasma kinetic pressure is not constant while instability occurs. Such an instability does not fit in the conventional entropy mode category, but might be better referred to as an electromagnetic universal instability. The dispersion relation derived in our previous work assumed singly charged ions and neglected electron finite Larmor radius (FLR) effects -- a common practice for analyzing ion-scale instabilities. In this paper, we extend our derivation to include arbitrary charge state, and full electron FLR and Debye shielding effects, and hence attain a more accurate generalized dispersion relation applicable to a wider range of plasmas. In particular, we examine solutions in an electron-positron (pair) plasma environment, where electron FLR effects strongly influence the overall plasma behavior. Notably, we find that the instability persists in this simple plasma system.

Electron-positron plasma research is an active area of inquiry in plasma physics and astrophysics. Though it is conceptually simple, producing a sufficient number of electron-positron pairs to form a plasma and studying its collective behavior before annihilation in the laboratory is challenging. Even though the original idea of a pair plasma experiment and the first theoretical study of its properties (e.g., the absence of Faraday rotation, ion acoustic and drift waves because of the exact mass asymmetry) dates back to the late 1970s,~\cite{tsytovich1978laboratory} only very recently have active experiments on pair plasma been proposed~\cite{chen2011towards,pedersen2012plans} and carried out.~\cite{chen2015scaling,saitoh2015efficient,warwick2017experimental} Two main ongoing pair plasma experiments include the A Positron-Electron Experiment (APEX) in Germany and the experiment in Lawrence Livermore National Laboratory (LLNL) in the United States. APEX produces non-relativistic positrons by the pair production process from absorption of MeV $\gamma$-radiation in platinum, and plans to trap (and neutralize) them in a magnetic dipole~\cite{hergenhahn2018progress}. The LLNL experiment generates a relativistic pair plasma via energetic short-pulse laser irradiation and plans to confine it in a magnetic mirror.~\cite{jens2019}
In part motivated by such experiments, theoretical and numerical investigations of instability and transport in pair plasmas have also surged in the past few years.
For example, gyrofluid~\cite{kendl2017interchange} and linear gyrokinetic~\cite{kennedy2018linear,kennedy2020linear} simulations were performed to study pair plasma's stability in dipole and tokamak/stellarator configurations.
Meanwhile, recent electrostatic~\cite{helander2014microstability} and electromagnetic (with incompressible $B_\parallel$)~\cite{helander2016gyrokinetic} work concludes that all microinstabilities are absent in pair plasmas with a homogeneous magnetic field.
Further studies suggest that under certain circumstances microinstabilities do exist in a slab pair plasma, e.g., current driven instabilities are supported in a sheared slab,~\cite{zocco2017slab} drift instabilities can reappear when ion impurities are present and the ion fraction exceeds some threshold,~\cite{mishchenko2018gyrokinetic}, alternatively density or temperature gradient driven instabilities could be excited in non-neutral electron-positron plasmas.~\cite{kennedy2019local} The analysis we performed in this paper allows $\delta B_\parallel\neq 0$, therefore, our finding that the $\delta B_\parallel$-universal instability can be unstable in pair plasmas does not contradict previous studies, but rather is an extension with a relaxed assumption on the guide field.

This paper is organized as follows. The generalized dispersion relation of the slab universal instability is derived in section~\ref{sec:gdr}.
Discussions concerning the impact of electron finite Larmor radius (FLR) and Debye shielding effects are presented in section~\ref{sec:flr}. 
In section~\ref{sec:ep} we show that this instability persists in a simple electron-positron pair plasma slab. Section~\ref{sec:conc} summarizes our key findings.

\section{Generalized dispersion relation} \label{sec:gdr}

Consider a quasi-neutral plasma consisting of electrons and one other species with positive charge (e.g., positrons or cations) in a two-dimensional $(xy)$ slab with a guiding magnetic field $\boldsymbol{B}(x)$ aligned with the $z$ direction. For simplicity we still refer to this positively charged species as ions with charge $q_i{=}Ze$ and mass ratio $\mu{=}m_i/m_e$. The equilibrium plasma pressure is then $p_0{=}p_{0i}{+}p_{0e}{=}n_{0i}T_{0i}(1{+}Z\tau_e)$ where the quasi-neutral condition $n_{0e}{=}Zn_{0i}$ has been applied and $\tau_e{=}T_{e0}/T_{i0}$ is the ratio between electron and ion temperatures. Note that the quasi-neutral condition also ensures $L_{n_i}{=}L_{n_e}{=}L_n$ where the characteristic gradient scale length for a quantity $f$ is $L_f=f/f'$, with $f'=\partial f/\partial x$. The equilibrium pressure balance condition $p_0{+}B_0^2/(8\pi){=}\text{constant}$, therefore, implies that $L_B^{-1}{=}-\beta L_p^{-1}/2$, or
\begin{equation}\label{eq:mhd_eq}
\frac{L_{n_i}}{L_B}=-\frac{\beta_i\alpha_0}{2},
\end{equation}
with
\begin{equation}
\alpha_0=1+\eta_i+Z\tau_e+Z\tau_e\eta_e,\
\end{equation}
$\beta_{\alpha}=8\pi p_{\alpha 0}/B_0^2$, and the characteristic length ratios are $\eta_\alpha=(n_0T_{\alpha 0}')/(n_0'T_{\alpha 0})$ for $\alpha=e,i$.
Enforcement of the equilibrium pressure balance condition within the slab geometry (i.e., when the magnetic tension force associated with magnetic curvature is negligible) is proven to be crucial in order to avoid a specious instability driven by the pressure gradients, even in low $\beta$ cases.~\cite{rogers2018gyrokinetic}

Defining the thermal speed $v_{t \alpha}=\sqrt{{2T_{\alpha 0}}/{m_\alpha}}$, the gyrofrequency $\omega_{c\alpha}=q_\alpha B_0/(m_\alpha c)$, and the gyroradius $\rho_\alpha=v_{t\alpha}/\omega_{c\alpha}$, assuming $k_\parallel=0$ and the perturbed magnetic field $\tilde{\boldsymbol{B}} \propto e^{-i\omega t+ik_\perp y} \hat{\boldsymbol{z}}$, expanding the linearized Vlasov equation and distribution functions according to the gyrokinetic ordering $\epsilon\sim\rho_\alpha/L$, and then substituting the perturbed electron and ion distribution function in Amp\`ere's law and Gauss's law, yields the generalized dispersion relation of the slab $\delta B_\parallel$-universal instability
\begin{equation}\label{eq:gdr}
I_{\phi Q} I_{BA}=I_{BQ}I_{\phi A}=2\left(I_{\phi A}\right)^2/(Z\beta_i)
\end{equation}
with
\begin{eqnarray}
I_{\phi Q}&=&2\int_0^\infty dv v e^{-v^2}\left[
Z\left(J_0^2\frac{\bar{\omega}_i}{\omega_{bi}}-1\right)+\left(J_0^2\frac{\bar{\omega}_e}{\omega_{be}}-1\right)\tau_i\right] - k^2\frac{\lambda_{De}^2}{\rho_i^2}\tau_i \ , \label{eq:IphiQt}\\
I_{BA}&=&1+4\int_0^\infty dv v^3 e^{-v^2}\left( J_1^2\frac{\beta_i}{k^2}\frac{\bar{\omega}_i}{\omega_{bi}}+\frac{\mu}{Z^2\tau_e}J_1^2\frac{\beta_e}{k^2}\frac{\bar{\omega}_e}{\omega_{be}}\right)\ ,\label{eq:IBAt}  \\
I_{\phi A}&=&-2Z\beta_i \int_0^\infty dv v^2 e^{-v^2}\left( J_0J_1\frac{1}{k}\frac{ \bar{\omega}_i}{\omega_{bi}} +\frac{\mu^{1/2}}{Z\tau_e^{1/2}}J_0J_1\frac{1}{k}\frac{ \bar{\omega}_e}{\omega_{be}}\right). \label{eq:IphiAt}
\end{eqnarray}
Here $\tau_i=T_{i0}/T_{e0}=\tau_e^{-1}$, $v=v_{\perp}/v_{t}$,
\begin{eqnarray}
\bar{\omega}_i=\omega-\frac{k}{2}\left[ 1+\eta_i \left(v^2-1\right)\right] \ , \quad \omega_{bi}= \omega+\frac{\beta_i\alpha_0}{4}kv^2 , \\
\bar{\omega}_e=\omega+\frac{Z\tau_e k}{2}\left[ 1+\eta_e \left(v^2-1\right)\right] \ , \quad \omega_{be}=\omega-\frac{\beta_e\alpha_0}{4}kv^2,
\end{eqnarray}
$J_0$ and $J_1$ are Bessel functions of the first kind, the arguments of $J_{0,1}$ for ions and electrons are $kv$ and $-Z\sqrt{\tau_e/\mu}\,
kv$ respectively, and $\omega,k$ are normalized according to
\begin{equation}
    \omega=\omega_\text{phys}L_n/v_{ti}, \quad k=k_\text{phys}\rho_i.
\end{equation}

A complete description of the derivation of the generalized dispersion relation is outlined in the supplementary material. In such derivation no assumption was made on charge state, full electron FLR effects were retained and the perturbed electric field was calculated through Gauss's law. 
As a result, the new dispersion relation is not only a function of $k_\perp$, plasma $\beta$, temperature ratios $\tau_{i,e}$ and characteristic length ratios $\eta_{i,e}=L_n/L_{T_{i,e}}$, but also depends on the charge state $Z$, the ion-electron mass ratio $\mu$ and the Debye length $\lambda_{D_e}$.

As expected, if one takes small argument expansions of Bessel functions 
\begin{equation}
    J_0(b){=}1{-}b^2/4{+}O(b^4),\quad J_1(b){=}b/2{-}b^3/16{+}O(b^5)
\end{equation}
and considers dense plasmas ($\lambda_{De}/\rho_i{\ll} 1$) with singly charged ions ($Z{=}1, \mu{=}m_i/m_e{\gg} 1$ so that only leading order FLR effects of electrons are kept), then the above equations recover the dispersion relation without higher order electron FLR and Debye shielding correction (i.e., equations (17)-(19) of reference~\cite{rogers2018gyrokinetic}).

\begin{figure}
	\centering
	\includegraphics[width=.8\textwidth]{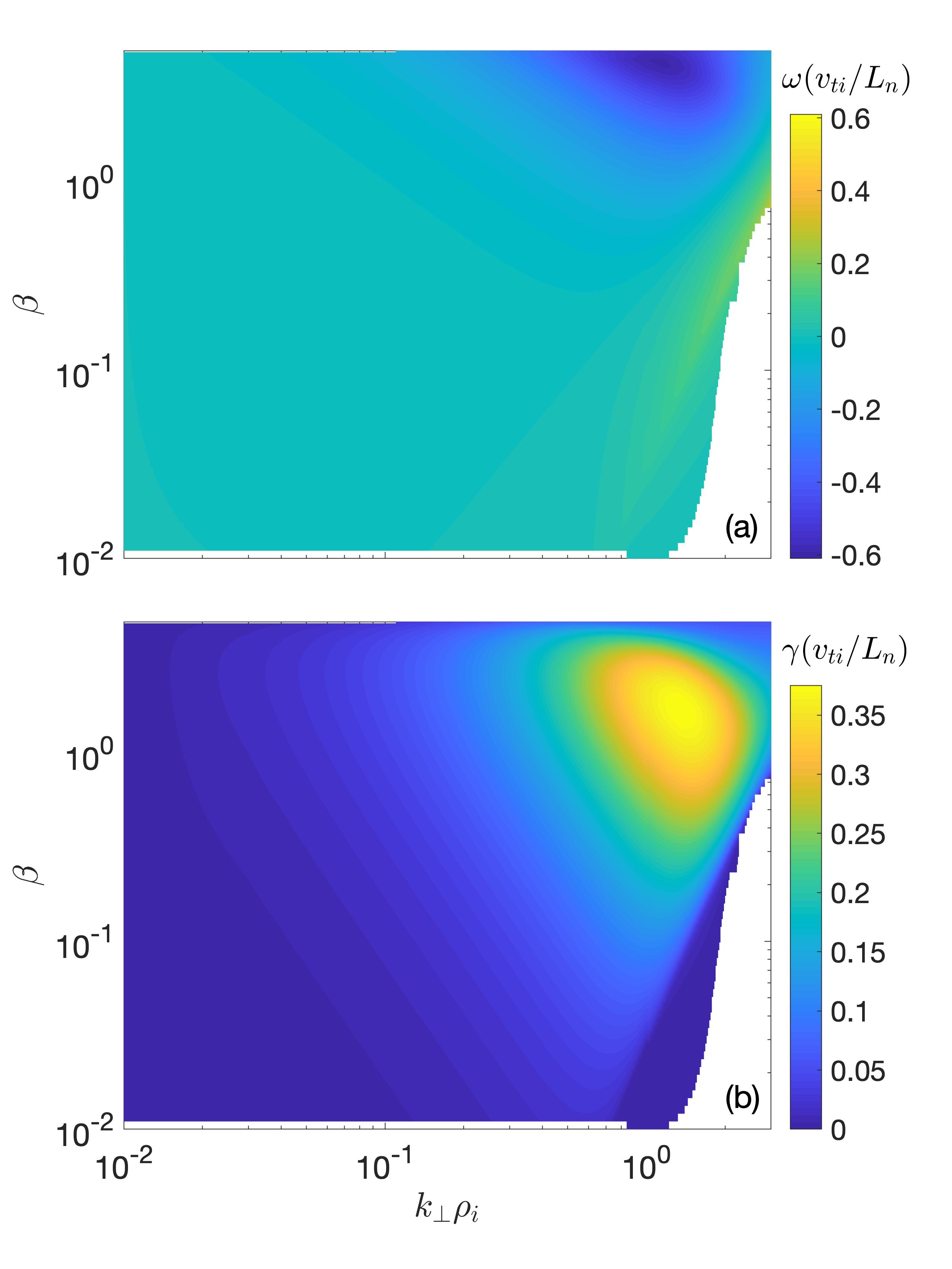}
	\caption{Slab universal mode (a) frequency $\omega$ and (b) linear growth rate $\gamma$ verses $k_\perp,\beta$ in hydrogen plasma ($\mu=1836,Z=1$) with $\tau_e=1,\eta_i=-1,\eta_e=2$ and $\lambda_{D_e}\approx0$.}
	\label{fig:lgr}
\end{figure}

Because the slab $\delta B_\parallel$-universal mode is also driven by density and temperature gradients, like conventional entropy modes, it shares characteristics of the entropy modes.
In general, the linear growth rate of the slab $\delta B_\parallel$-universal mode vanishes in the long wavelength $k\rho_i\ll 1$ limit and peaks near $k\rho_i\sim 1$. This mode tends to propagate in the electron diamagnetic direction at low $k$ while it reverses to the ion diamagnetic direction at high $k$ (the real frequency $\omega$ changes from negative to positive) as elucidated in figure~\ref{fig:lgr}. In addition, this mode is somewhat ``universal" -- for any slab plasma with fixed $\beta$ and varying guide magnetic field (i.e., $L_B\neq \infty$ or equivalently the plasma pressure $p$ is inhomogeneous), there is always a constrained parameter region (negative $\eta_i$ and/or $\eta_e$) in which this mode is unstable.

\begin{figure}
	\centering
	\includegraphics[width=.8\textwidth]{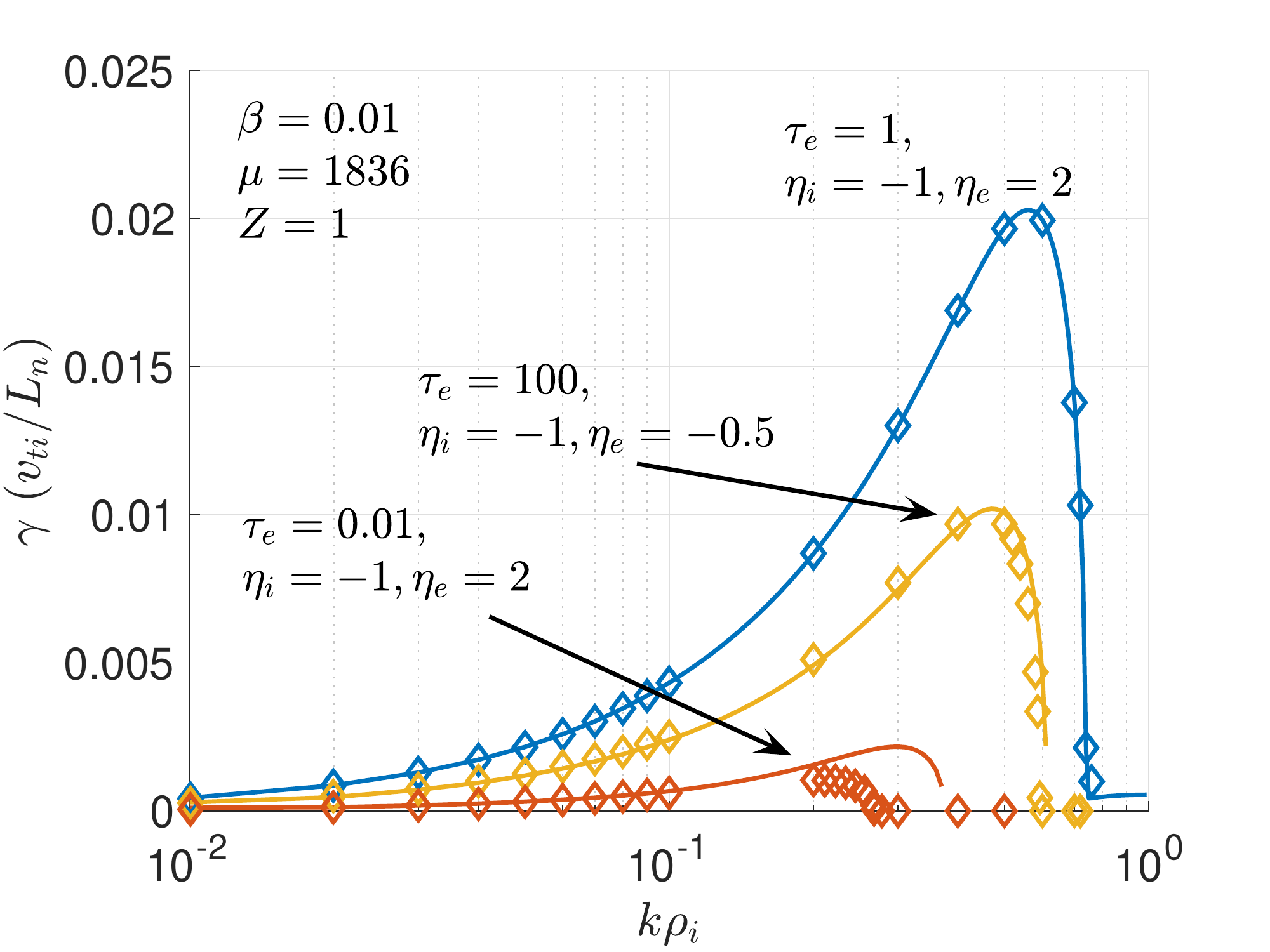}
	\caption{Linear growth rate of hydrogen plasma from general dispersion relation (solid lines) along with result from GENE eigenvalue solver (diamond markers).}
	\label{fig:gene}
\end{figure}

Without further simplification, the generalized dispersion relation of the slab universal instability (equation~\ref{eq:gdr}) is too complicated to be solved analytically. We therefore solve it using a numerical root-finding algorithm. In order to validate the generalized linear dispersion relation and verify that it is solved correctly, the numerical solutions of equation~\ref{eq:gdr} are benchmarked with results of GENE's linear eigenvalue solver.~\cite{jenko2000electron} As shown in figure~\ref{fig:gene}, overall good agreement has been achieved over a wide range of parameters; a mild discrepancy only appears at moderate $k$ for low $\beta$ and extremely hot ions (red line and diamonds in figure~\ref{fig:gene}).

%\subsection{low $\beta$ and long wavelength limit}
Some analytical progress can be made, however, in simpler limiting cases. In the low $\beta$ and long wavelength limit, an analytical dispersion relation is derived by taking the small argument approximations of Bessel functions $J_0$ and $J_1$.
Defining the normalized phase velocity $u=2\omega/k$ and the normalized Debye length $\bar{\lambda}_{D_e}^2=2\lambda_{D_e}^2\tau_i/(Z\rho_i^2)$, to $O(\beta)$, $O(Z)$ and $O(k^2)$, the dispersion relation becomes the quadratic equation
\begin{equation}\label{eq:lowkdr}
    a_2u^2+a_1u+a_0=0
\end{equation}
with
\begin{eqnarray}
a_2&=&\left(1+\frac{1}{\mu}+\bar{\lambda}_{D_e}^2\right)\left(1+\beta\right)=\left(1+\frac{1}{\mu}+\bar{\lambda}_{D_e}^2\right)\left[1+\beta_i\left(1+Z\tau_e \right) \right]\ , \\
a_1&=&-\alpha_{1i}\left(1+\beta\right)+\beta_i\left( \frac{\alpha_0}{2}-\alpha_4-\bar{\lambda}_{D_e}^2\alpha_4 \right) +\frac{1}{\mu}\left[ Z\tau_e\alpha_{1e}\left(1+\beta\right)-\beta_i\left( \frac{Z\tau_e\alpha_0}{2}+\alpha_4 \right)\right]\ , \\
a_0&=&\beta_i\left[ \alpha_{1i}\alpha_4-\frac{\alpha_0\alpha_{2i}}{2}-\frac{Z\tau_e}{\mu}\left( \alpha_{1e}\alpha_4+\frac{Z\tau_e\alpha_0\alpha_{2e}}{2}\right) \right]\ .
\end{eqnarray}
and
\begin{equation}
\alpha_{1i,e}=1+\eta_{i,e}\ , \quad \alpha_{2i,e}=1+2\eta_{i,e}\ , \quad  \alpha_4=1-Z^2\tau_e^2+2\left(\eta_i-Z^2\tau_e^2\eta_e\right),
\end{equation}
The instability condition thus becomes
\begin{equation}
    \Delta=a_1^2-4a_0a_2<0
\end{equation}
and the linear growth rate is
\begin{equation}\label{eq:lowkgamma}
    \gamma_\text{est}=\text{Im}(\omega)=\frac{|\Delta|^{1/2}k}{4a_2}.
\end{equation}

\begin{figure}
	\centering
	\includegraphics[width=.8\textwidth]{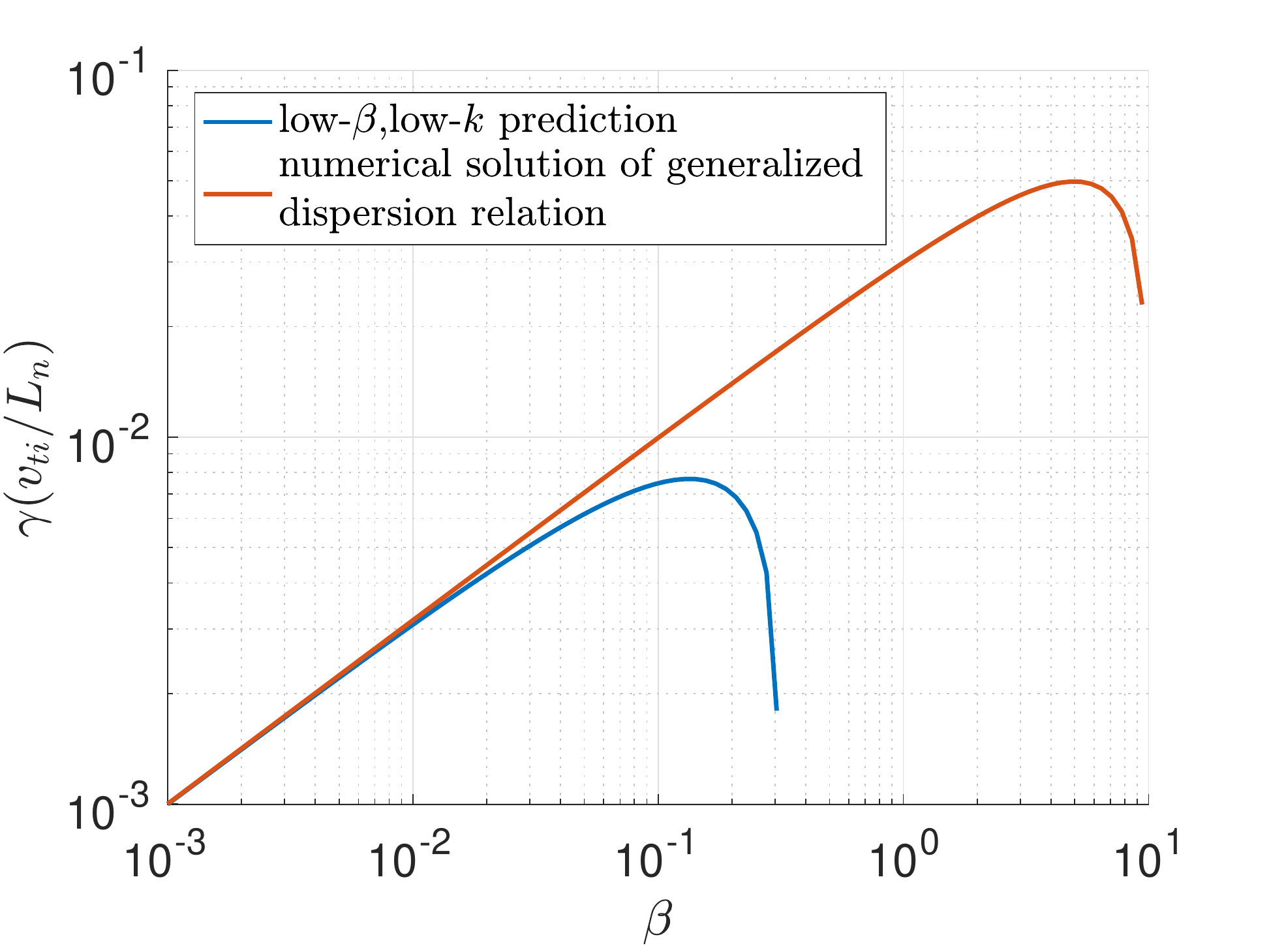}
	\caption{Numerical (red) and analytical (blue) growth rates in singly charged neon plasma for $k_\perp\rho_i=0.1,\tau_e=2,\eta_i=-1,\eta_e=2$ and $\lambda_{D_e}=0$.}
	\label{fig:lowbeta}
\end{figure}

Details of the derivation are outlined in appendix~\ref{app:lowbeta}. As plotted in figure~\ref{fig:lowbeta}, compared to the numerical solution of full dispersion relation in equation~\ref{eq:gdr}, the low $\beta$ long wavelength estimate in equation~\ref{eq:lowkgamma} typically yields an accurate linear growth rate in the low $\beta$ regime ($\beta \leq 0.1$) but fails to predict instability as $\beta$ approaches unity. 
Nevertheless, it is still a very useful formula as it retains electron FLR and Debye shielding effects and is applicable to arbitrary $\tau_{i,e}$. Therefore, we will use it to explore the stability conditions, quantify the impact of electron FLR effects and help us understand instability properties in the low $\beta$ long wavelength regime.

\section{Electron FLR and Debye shielding effects}\label{sec:flr}
In the gyrokinetic study of ion-scale instabilities, a common means of simplifying calculations is to drop high order electron FLR effects because they are normally two orders of magnitude smaller than ion FLR effects. Although recent multi-scale nonlinear simulations~\cite{maeyama2015cross,howard2015multi} show that electron scale instability can influence ion scale transport due to cross-scale coupling, neglecting electron FLR effects in the linear analysis of ion-scale modes is still considered justified, at least for electron-ion plasmas.
In this section, we explore the impact of electron FLR effects on the slab $\delta B_\parallel$-universal instability within the linear gyrokinetic analysis, especially when the electron gyro-radius is no longer too small to be neglected.
We define the ratio of the electron and ion gyro-radius $\delta=|\rho_e/\rho_i|=Z\sqrt{\tau_e/\mu}$ as a (rough) indicator of the relative influence of electron and ion FLR effects.
Typically, $\delta$ is much smaller than unity. For example, $\delta \approx0.02$ for hydrogen plasma with $T_i\sim T_e$, and $\delta \approx0.05$ for fully ionized neon plasma with $T_i\sim T_e$. 

\begin{figure}
	\centering
	\includegraphics[width=.8\textwidth]{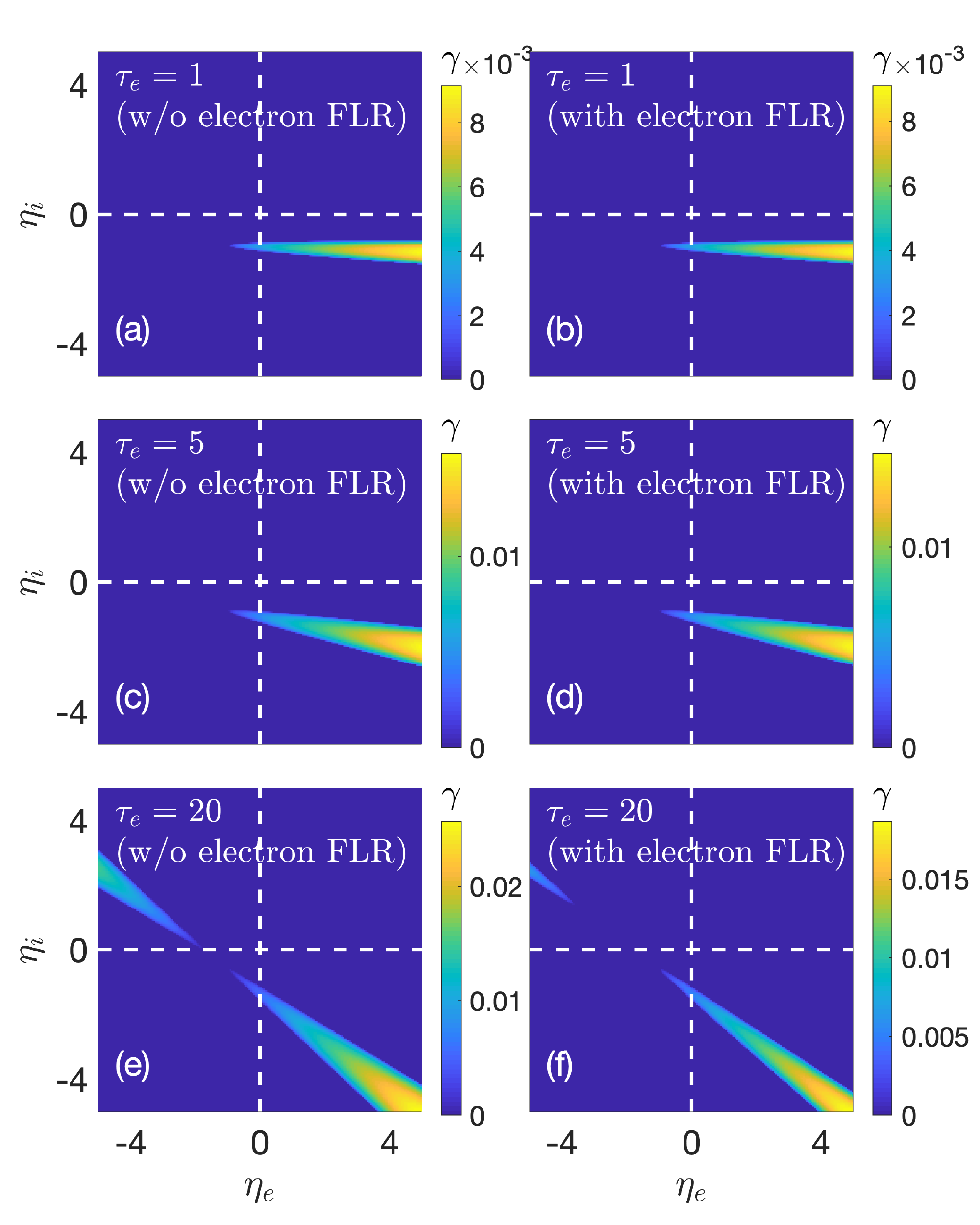}
	\caption{Slab universal instability linear growth rate as a function of $\eta_e,\eta_i,\tau$ for hydrogen plasma ($\mu=1836,Z=1$) at $k_\perp\rho_i=0.1,\beta=0.02,\lambda_{D_e}=0$ with and without electron FLR correction.}
	\label{fig:eflr_tau1}
\end{figure}

We first quantify the corrections to our previous analysis of slab $\delta B_\parallel$-universal instability (without electron FLR effects)~\cite{rogers2018gyrokinetic} due to electron FLR effects.
Figure~\ref{fig:eflr_tau1} compares the instability conditions and linear growth rates of the slab $\delta B_\parallel$-universal instability based on the low $\beta$ long wavelength linear growth rate expression~\ref{eq:lowkdr} with and without electron FLR effects for a hydrogen plasma at $\beta=0.02,k_\perp\rho_i=0.1$ and $\lambda_{D_e}\approx0$.
As $\delta\propto\sqrt{\tau_e}$, equal temperature ($\tau_e=1$), hot electron ($\tau_e=5$) and extremely hot electron $\tau_e=20$ cases are chosen.
In the equal temperature ($\tau_e=1,\delta=0.02$) and hot electron ($\tau_e=5,\delta=0.05$) plasmas, the electron FLR correction is too small to be distinguished, indicating that neglecting the electron FLR effects is indeed justified for these cases.
For the extremely hot electron case ($\tau_e=20,\delta=0.1$), including electron FLR effects has a stabilizing effect on the instability. For this very hot electron system the unstable region in $\eta_{e,i}$ space shrinks slightly and the maximum linear growth rate decreases by $25\%$, yet the overall shape remains qualitatively the same.
Therefore, we conclude that the additional electron FLR effects do not introduce any qualitative change in the properties of the slab $\delta B_\parallel$-universal instability, at least for the electron-ion plasma parameters; and the findings based on our previous analysis~\cite{rogers2018gyrokinetic} are still valid.
It is important to point out that this instability occurs in a parameter regime with $\eta_i<0$ and/or $\eta_e<0$, i.e., density and temperature gradients are in opposite directions. Although less common, negative $\eta$ could exist, for example, in a disrupted tokamak plasma. It also can be deliberately attained by locally heating electrons and/or ions and fueling in a laboratory setting.

\begin{figure}
	\centering
	\includegraphics[width=.8\textwidth]{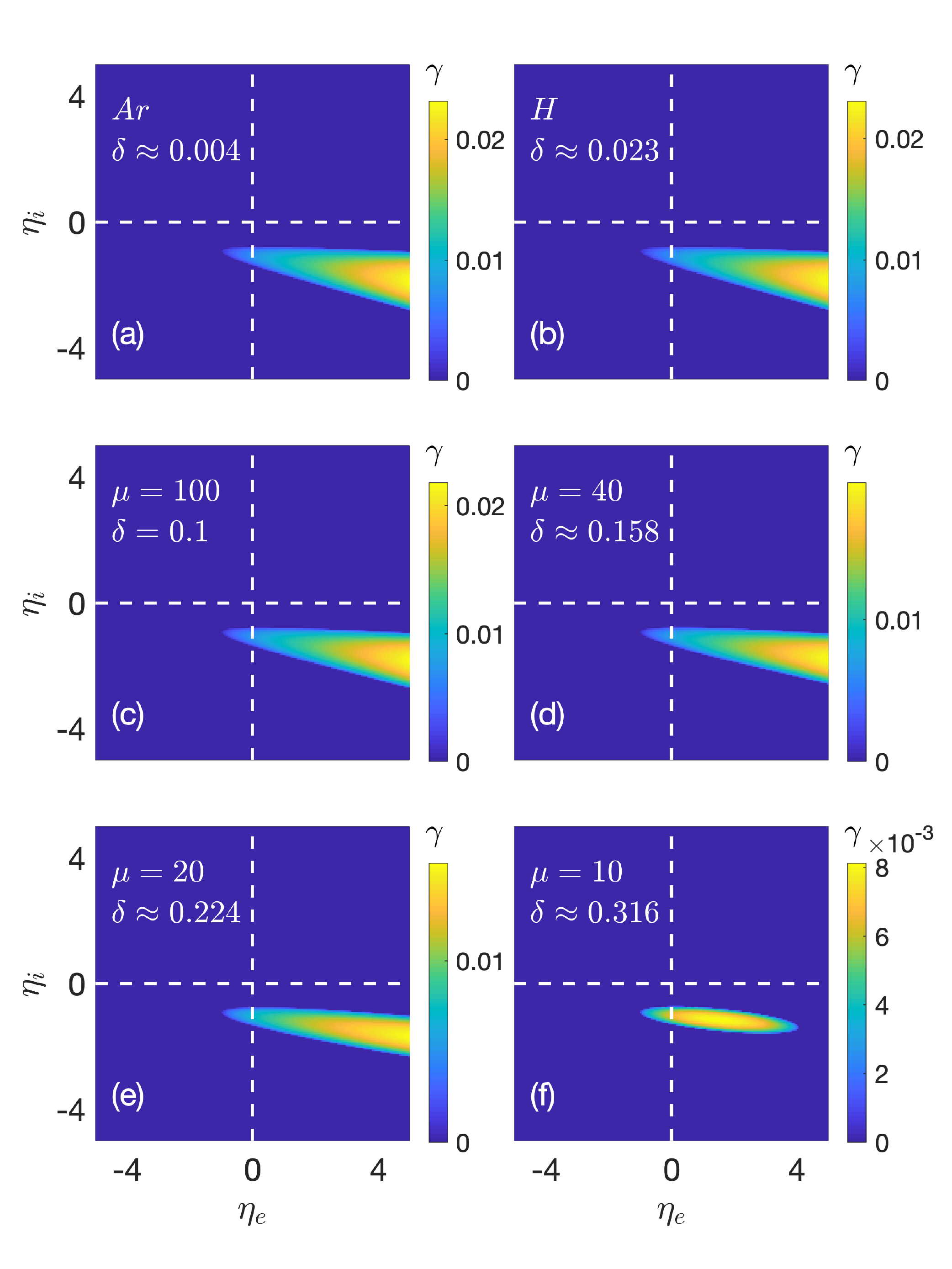}
	\caption{Slab universal instability linear growth rate as a function of $\eta_e,\eta_i,\mu$ for $k_\perp\rho_i=0.1,\beta=0.1,\tau_e=1,\lambda_{D_e}=0$.}
	\label{fig:eflr_mu1}
\end{figure}

\begin{figure}
	\centering
	\includegraphics[width=.8\textwidth]{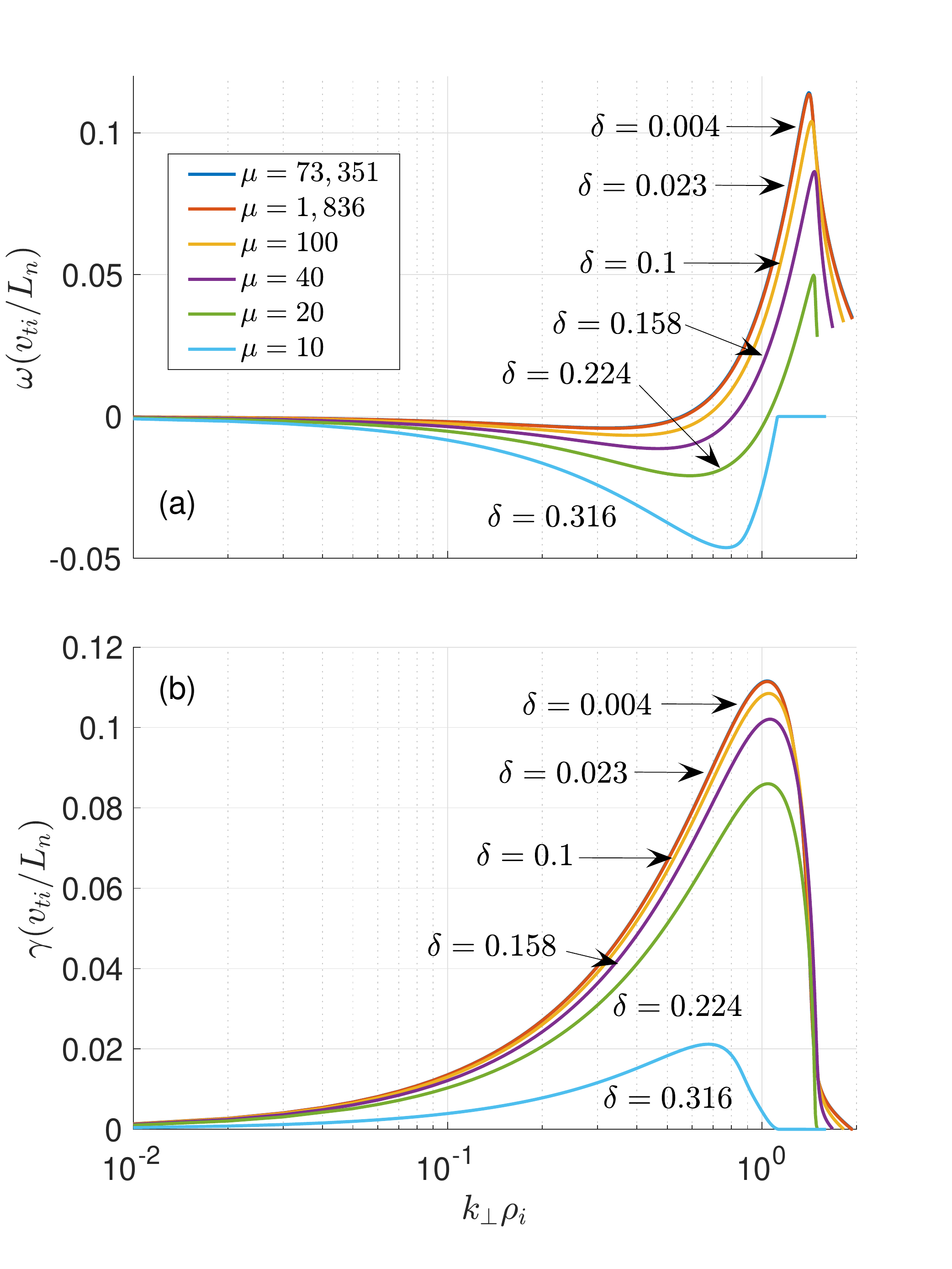}
	\caption{(a) Real frequencies and (b) linear growth rates of the slab universal instability for singly charged plasma ($Z=1$) with various ion mass and $\beta=0.1,\tau_e=1,\eta_e=2,\eta_i=-1,\lambda_{D_e}=0$.}
	\label{fig:eflr_mu2}
\end{figure}

In order to further investigate the impact of electron FLR effects, stability analyses are performed for six different types of plasmas with different ion species. We now consider singly charged argon plasmas ($\mu=73,351$), hydrogen plasmas ($\mu=1836$) and plasmas with artificially reduced ion-electron mass ratios $\mu\in\{100,40,20,10\}$, with fixed total plasma $\beta=0.1,\tau_e=1,k_\perp\rho_i=0.1$ and $\lambda_{D_e}\approx 0$. These plasmas hence correspond to $\delta\in\{0.004,0.023,0.1,0.158,0.224,0.316\}$ respectively.
Figure~\ref{fig:eflr_mu1} shows how the instability condition and linear growth rate evolve as the ion mass decreases, or $\delta$ increases. The stabilizing effect due the electron FLR contribution is more profound in this mass scan. For $\delta \leq 0.1$ (singly charged argon and hydrogen plasmas), the unstable region in $\eta_{e,i}$ space barely changes and the peak linear growth rates remains the same. As $\delta$ continues increasing to sub-unity ($\sim 0.3$), electron FLR effects start to have substantial influence on the instability: the unstable region becomes more constrained in $\eta_{e,i}$ space and the maximum linear growth rate reduces to roughly $1/3$ of its original value -- both indicate that the slab $\delta B_\parallel$-universal instability is stabilized by the electron FLR effects. This conclusion is further verified by solving the generalized dispersion relation in equation~\ref{eq:gdr} for plasmas of different ion mass $\mu$ at $Z=1,\beta=0.1,\tau_e=1,\eta_e=2,\eta_i=-1$ and $\lambda_{D_e}\approx 0$. Figure~\ref{fig:eflr_mu2} demonstrates that as the electron FLR effects increase, not only do the linear growth rates diminish, but also the mode  favors travel in the electron direction (negative in our notation) at low $k$, and the propagation transition point (i.e., $\omega=0$) is pushed towards higher $k$.

\begin{figure}
	\centering
	\includegraphics[width=.8\textwidth]{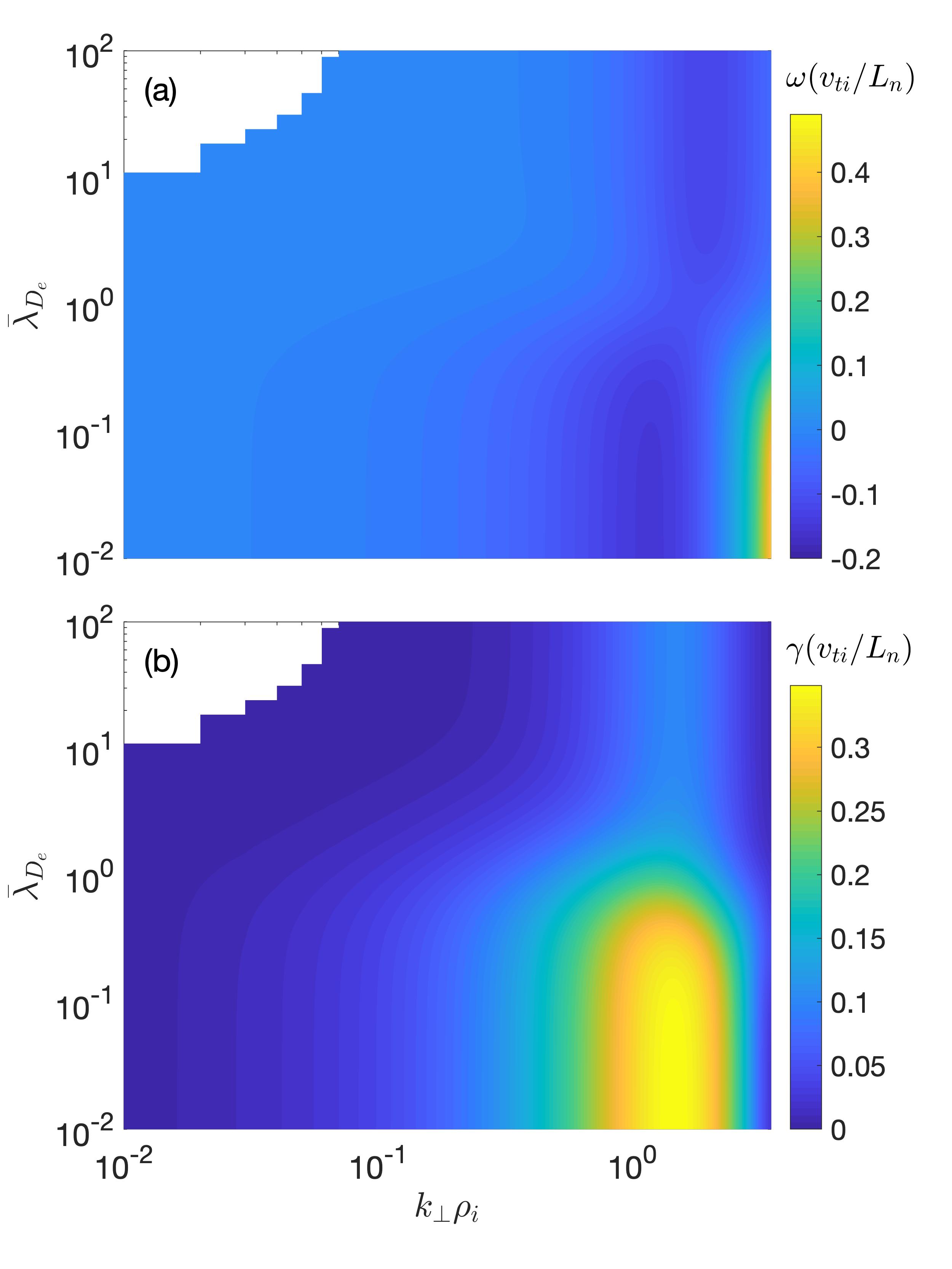}
	\caption{Stabilization of the slab universal instability for hydrogen plasma with $\beta=1,\tau_e=1,\eta_i=-1$ and $\eta_e=2$ as normalized Debye length $\bar{\lambda}_{D_e}=\lambda_{D_e}/\rho_i$ increases.}
	\label{fig:klambda_scan}
\end{figure}

The impact of Debye shielding on the slab universal instability is also studied. Not surprisingly, it is always stabilizing and starts to play an important role once $\lambda_{D_e}$ approaches $\rho_i$ as shown in figure~\ref{fig:klambda_scan}. 
Physically, the slab $\delta B_\parallel$-universal mode discussed in this paper is an electromagnetic instability -- perturbed electron and ion distribution functions produce a perturbed electrostatic potential $\tilde{\phi}$ as well as a perturbed current $\tilde{j}$; also, the characteristic wavelength of this mode $\lambda\sim 1/k_\perp\sim \rho_i$.
On the other hand, by definition, the Debye length $\lambda_{D}$ measures how far the electrostatic effects can extend in a plasma -- with each distance of $\lambda_D$, the effective electric potential decreases by a factor of $1/e$. 
If $\lambda_D \ll \lambda$, Debye shielding has little influence on $\tilde{\phi}$; however, as the Debye length increases towards $\lambda$, the electrostatic potential (and its perturbation) will be attenuated more and more strongly; eventually, when $\lambda_D \gg \lambda$, $\tilde{\phi}$ will be completely screened.

The Debye shielding effect can also be understood algebraically with the aid of the low $\beta$ long wavelength analytical formula in equation~\ref{eq:lowkdr}. If one neglects electron FLR effects ($\mu\to\infty$), then $a_2\approx (1+\bar{\lambda}_{D_e}^2)(1+\beta)$, $a_1 \approx -\alpha_{1i}(1+\beta)+\beta_i(\alpha_0/2-\alpha_4-\bar{\lambda}_{D_e}^2\alpha_4)$, and the instability condition with $\bar{\lambda}_{D_e}$ ordering yields
\begin{equation}
    \Delta\sim  \beta_i^2\alpha_4^2\bar{\lambda}_{D_e}^4 -O(\bar{\lambda}_{D_e}^2)<0.
\end{equation}
Clearly, given a large enough $\bar{\lambda}_{D_e}$, the $O(\bar{\lambda}_{D_e}^4)$ term will dominate and $\Delta>0$. In other words, the slab universal instability will be completely stabilized if the plasma density is sufficiently low such that $\bar{\lambda}_{D_e}\gg 1$ or $\lambda_{D_e}\gg\rho_i$. 
However, in many space and laboratory magnetized plasmas, $\rho_i\gg\lambda_{D_e}$; e.g., for solar wind plasma $\rho_i\sim 5\times10^4 \text{m}\gg \lambda_D\sim 10\text{m}$, for LAPD plasma $\rho_i\sim 2\times10^{-3} \text{m}\gg \lambda_D\sim 1\times10^{-5}\text{m}$, and for tokamak plasma $\rho_i\sim 2\times10^{-3} \text{m}\gg \lambda_D\sim 2\times10^{-5}\text{m}$. Therefore Debye shielding stabilization is expected to be weak in these plasma systems.

\section{Slab $\delta B_\parallel$-universal instability in pair plasma}\label{sec:ep}
In an electron-ion plasma, $\mu=m_i/m_e\gg 1$ and $T_e\sim T_i$ so that $\delta\ll 1$ and neglecting higher order electron FLR effects is justified as discussed in section~\ref{sec:flr}. However, this is not the case for electron-positron pair plasmas. 
Due to its unique mass symmetry feature ($m_i=m_p=m_e$ where subscript $p$ refers to positron), this simple plasma system has some fascinating properties. For example, large temperature separation is not expected in pair plasma as the ratios $\tau_{ee}:\tau_{pp}:\tau_{ep}=1:(m_p/m_e)^1/2:m_p/m_e=1:1:1$ where $\tau_{rs}$ is the collision time between species $r$ and species $s$.~\cite{iwamoto1993collective} Therefore, in a pair plasma, $\delta=\sqrt{\tau_e}\sim 1$; in other words, electron FLR effects are almost always equally as important as the ion (or, positron) FLR effects in a pair plasma.
Another interesting consequence of mass symmetry is that many instabilities (e.g., the electrostatic drift instability) will be absent in such a system.~\cite{tsytovich1978laboratory}

\begin{figure}
	\centering
	\includegraphics[width=.8\textwidth]{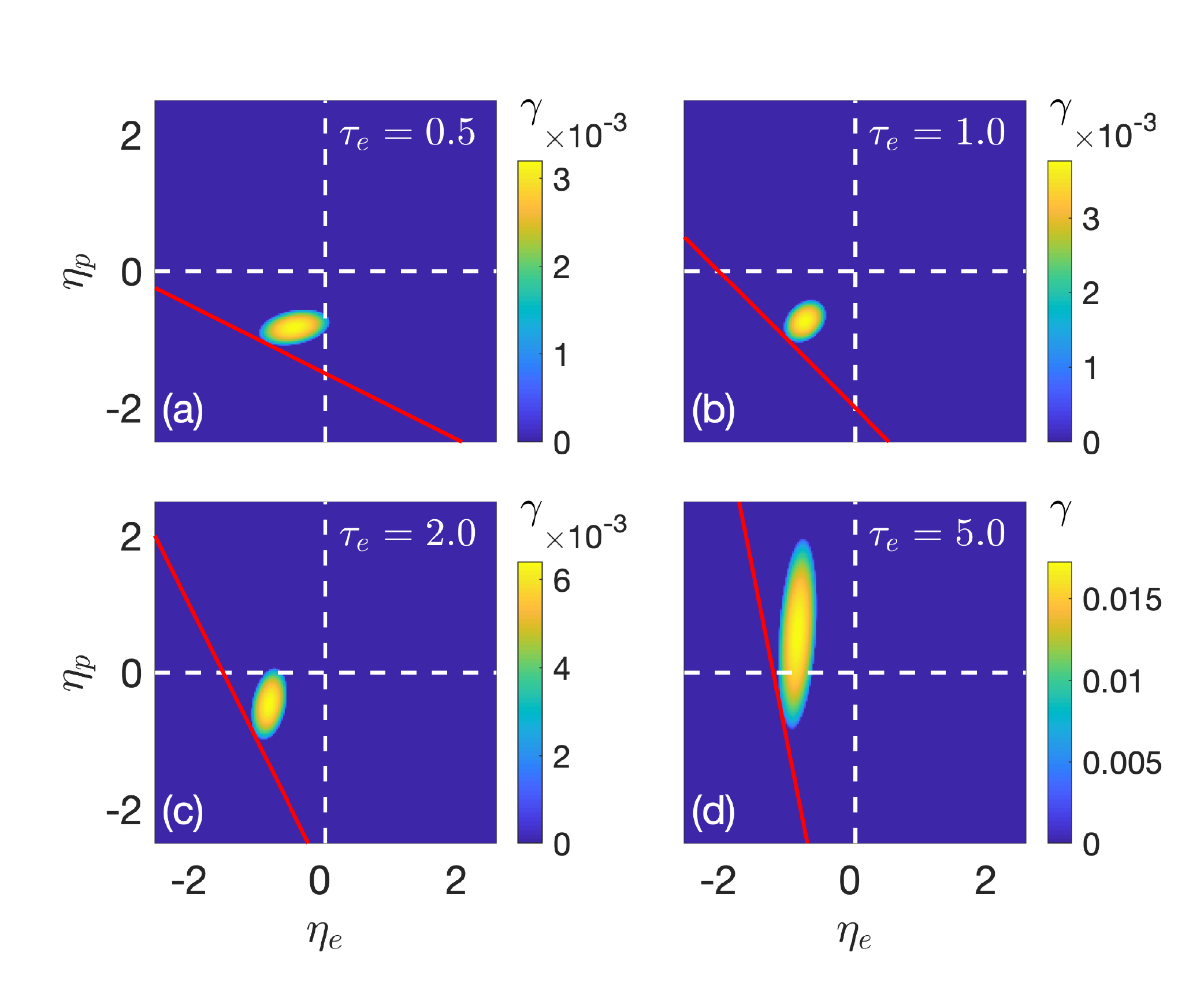}
	\caption{Slab universal instability linear growth rate as a function of $\eta_e,\eta_i,\tau$ for electron-positron plasma ($\mu=1,Z=1$) at $k_\perp\rho_i=0.1,\beta=0.1$. Red line denotes $L_p^{-1}=0$, i.e., plasma pressure is uniform.}
	\label{fig:ep_tau1}
\end{figure}

As shown in figure~\ref{fig:eflr_mu1}, the slab $\delta B_\parallel$-universal instability is stabilized when $\mu$ decreases (i.e., electron FLR effects become more important) -- the unstable region in $\eta$ space shrinks and the peak linear growth rate drops. One therefore might expect that this instability could be completely suppressed as $\mu$ is further decreased. However, we find that though suppressed substantially, the slab universal instability still exists when $\mu=1$ (figure~\ref{fig:ep_tau1}). In fact, the electron FLR effects alone cannot completely stabilize the slab $\delta B_\parallel$-universal instability in a pair plasma: when $\delta=|\rho_e/\rho_p|=\sqrt{\tau_e}<1$, the electron FLR effects tend to stabilize the instability; while for $\delta>1$ the electron FLR has a destabilizing effect on the instability.

We therefore conclude that the slab $\delta B_\parallel$-universal instability due to plasma inhomogeneity persists in the pair plasma. Our result, however, does not contradict the previous finding that ``there are no linear gyrokinetic instabilities in a pair plasma embedded in a constant magnetic field, regardless of the size of the density and temperature gradients."~\cite{helander2016gyrokinetic} This is because (1) this instability only exists when the magnetic field is not uniform (i.e., no intersection with the red lines in figure~\ref{fig:ep_tau1}); (2) we assume here that the guiding magnetic field is compressible ($\delta B_\parallel \neq 0$) instead of incompressible ($\delta B_\parallel=0$).

For electron positron pair plasma ($\mu=Z=1$), the general dispersion relation in equation~\ref{eq:gdr} becomes
\begin{equation}\label{eq:ep_gdr}
I_{\phi Q} I_{BA}=(2/\beta_p)\left(I_{\phi A}\right)^2
\end{equation}
with
\begin{eqnarray}
\begin{aligned}
I_{BA}&=1+4\int_0^\infty d\bar{v}_{\perp} \bar{v}_{\perp}^3 e^{-\bar{v}_{\perp}^2}\left( J_1^2\frac{\beta_p \bar{\omega}_p}{k^2\rho_p^2\omega_{bp}}+J_1^2\frac{\beta_e \bar{\omega}_e}{k^2\rho_e^2\omega_{be}}\right)\ , \\
I_{\phi Q}&=2\int_0^\infty d\bar{v}_{\perp} \bar{v}_{\perp} e^{-\bar{v}_{\perp}^2}\left[
J_0^2\frac{\bar{\omega}_p}{\omega_{bp}}-1-\left(J_0^2\frac{\bar{\omega}_e}{\omega_{be}}-1\right)\frac{T_{p0}}{T_{e0}}
\right]\ ,\\
I_{\phi A}&=-2\beta_p \int_0^\infty d\bar{v}_{\perp} \bar{v}_{\perp}^2 e^{-\bar{v}_{\perp}^2}\left( J_0J_1\frac{ \bar{\omega}_p}{k\rho_p \omega_{bp}} +J_0J_1\frac{ \bar{\omega}_e}{k\rho_e \omega_{be}}\right)\ .
\end{aligned}
\end{eqnarray}
Here
\begin{equation}
\bar{\omega}_p=\omega-(k/2)\left[1+\eta_p\left(v^2-1\right)\right] \ ,\quad \bar{\omega}_e=\omega+(\tau_e k/2)\left[1+\eta_e\left(v^2-1\right)\right]\ ,
\end{equation}
and
\begin{equation}
\omega_{bp}=\omega-(k/2)(L_n/L_B)v^2 \ ,\quad \omega_{be}=\omega+(\tau_e k/2)(L_n/L_B)v^2\ .
\end{equation}

\begin{figure}
	\centering
	\includegraphics[width=.8\textwidth]{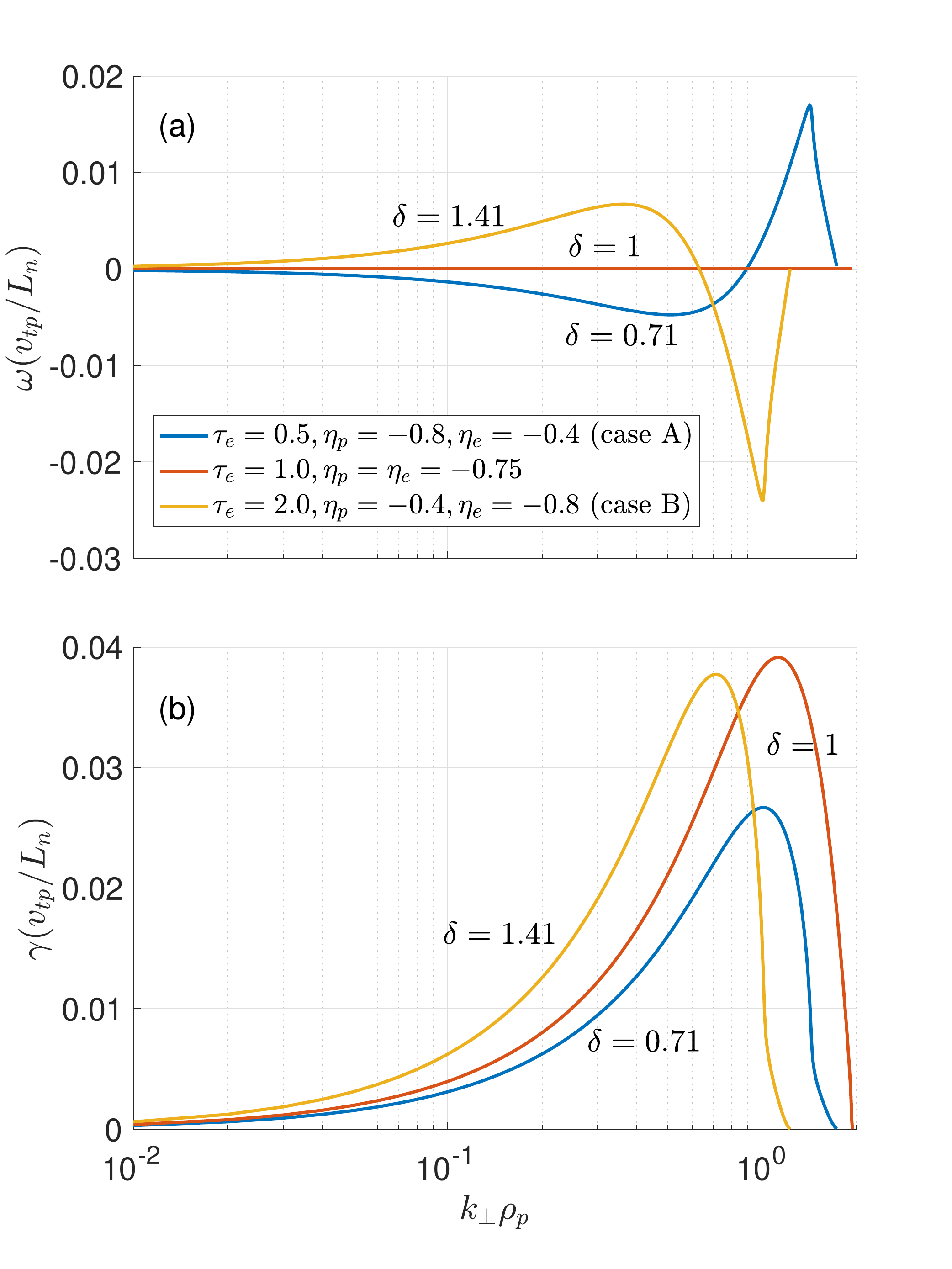}
	\caption{(a) Real frequencies and (b) linear growth rates of the slab universal instability in a pair plasma with $\beta=0.1$.}
	\label{fig:ep_tau2}
\end{figure}

To highlight the characteristics of this slab universal mode in a pair plasma, we solve the general dispersion relation of equation~\ref{eq:ep_gdr} and show, in figure~\ref{fig:ep_tau2}, the frequencies and growth rates as a function of $k_\perp\rho_p$ for three different (yet somewhat symmetric) sets of parameters. The self-symmetric setting (i.e., electrons and positrons have the same temperature profile as $\tau_e=T_e/T_p=1,\eta_i=\eta_e=-0.75$) has balanced electron and positron FLR effects ($\delta=1$), and results in a \textit{purely growing} instability, i.e., the real mode frequency $\omega$ is zero. This can be understood via the low-$\beta$, long wavelength analytical formula in section~\ref{sec:gdr}. With these self-symmetric parameters, the linear coefficient $a_1$ in the quadratic equation vanishes (assume $\bar{\lambda}_{D_e}=0$), and the solutions become two purely imaginary numbers corresponding to a decaying mode and an instability. 

\begin{figure}
	\centering
	\includegraphics[width=.8\textwidth]{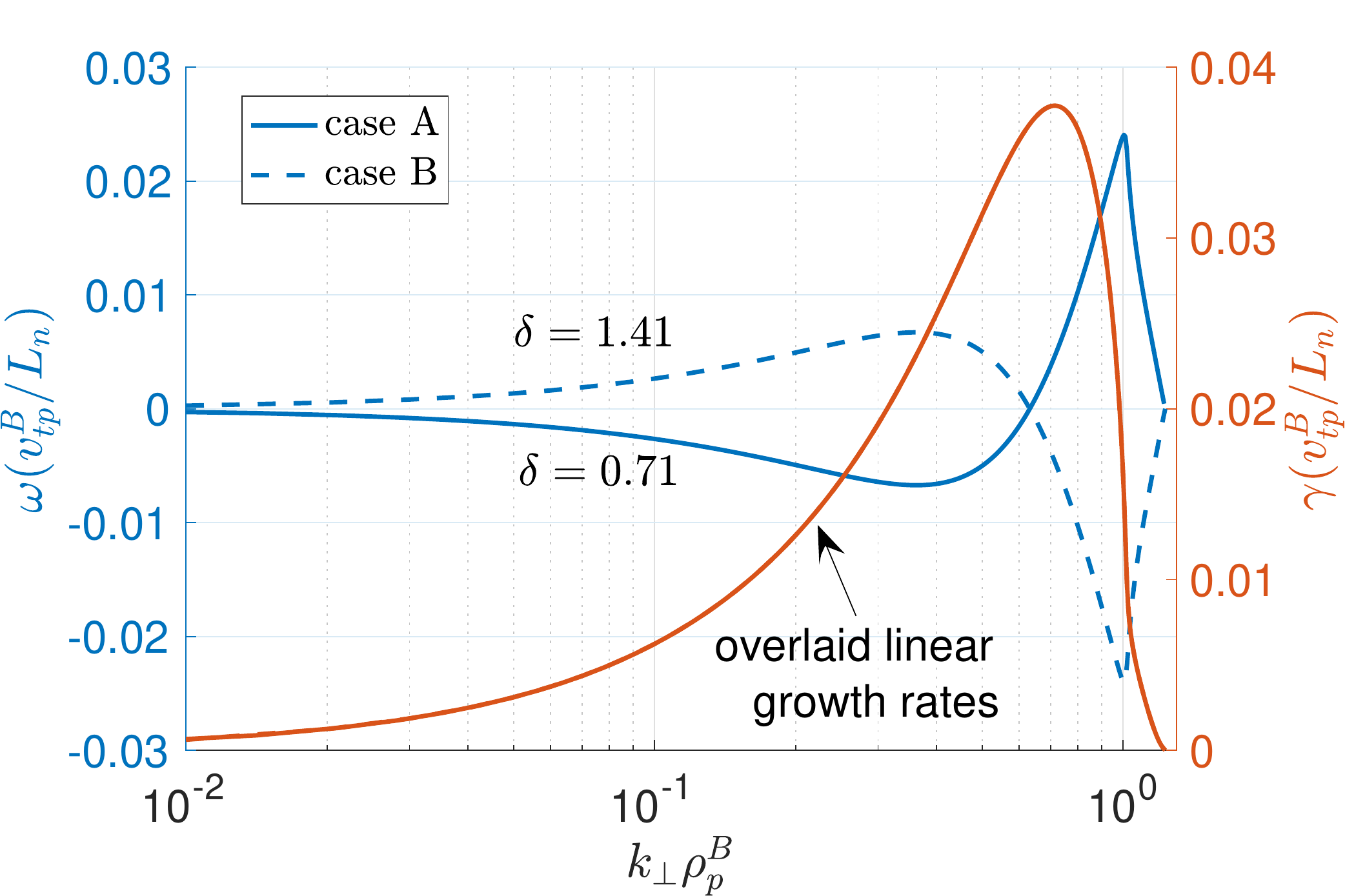}
	\caption{Real frequencies (blue) and linear growth rates (red) of the slab universal instabilities in a pair plasma with $\beta=0.1$ for case A (solid lines) and B (dashed lines) normalized with respect to case B positron/ion temperature.}
	\label{fig:ep_tau3}
\end{figure}

The mirror settings (case A: $\tau_e=0.5,\eta_p=-0.8,\eta_e=-0.4$ and case B: $\tau_e=2.,\eta_p=-0.4,\eta_e=-0.8$) reflect temperature profiles between electrons and positrons, and produce two anti-symmetric real frequency $k-$spectra and two identical linear growth rate curves (in physical units).
In figure~\ref{fig:ep_tau2}, the anti-symmetric and identical dispersion relations are less obvious because they are plotted in normalized units with respect to case-specific parameters. Recall that for a fixed $\beta$, the positron temperature in case A ($\tau_e=0.5$) is twice as much as in case B ($\tau_e=2$) as $T_p\propto\beta_p=\beta/(1+\tau_e)$, resulting in a $\sqrt{2}$ larger gyro-radius $\rho_p$ and thermal velocity $v_{tp}$ and hence the differences in figure~\ref{fig:ep_tau2} (i.e., curves of case A are stretched along $x$ axis and compressed along $y$ axis by $\sqrt{2}$ compared to the curves of case B). Take for example the most unstable point of two dispersion relations -- case A at $k_\perp\rho_p^A=1.01,\gamma_\text{max}/(v_{ti}^A L_n)=0.0267$ and case B at $k_\perp\rho_p^B=0.71,\gamma_\text{max}/(v_{ti}^B L_n)=0.0377$, are overlaid with each other in physical units or in normalized units with a fixed positron/ion temperature (as shown in figure~\ref{fig:ep_tau3}).

It is interesting to see that the mode propagation direction completely reverses once the electron FLR effects exceeds positron FLR effects (i.e., $\delta>1$) as indicated in figure~\ref{fig:ep_tau2}(a).
This can be understood by considering a slab pair plasma with species gyrating around a guiding magnetic field unstable to the slab $\delta B_\parallel$-universal instability. The excitation of this instability requires a non-uniform magnetic field but does not depend on the magnetic field direction. If electrons and positrons are suddenly swapped and the guiding field direction is simultaneously flipped, the new system has virtually the same assembly of particles gyrating in the same direction; one would thus expect the same instability with the mode propagating in the opposite direction with respect to the guiding field direction (as it is flipped).

\begin{figure}
	\centering
	\includegraphics[width=.8\textwidth]{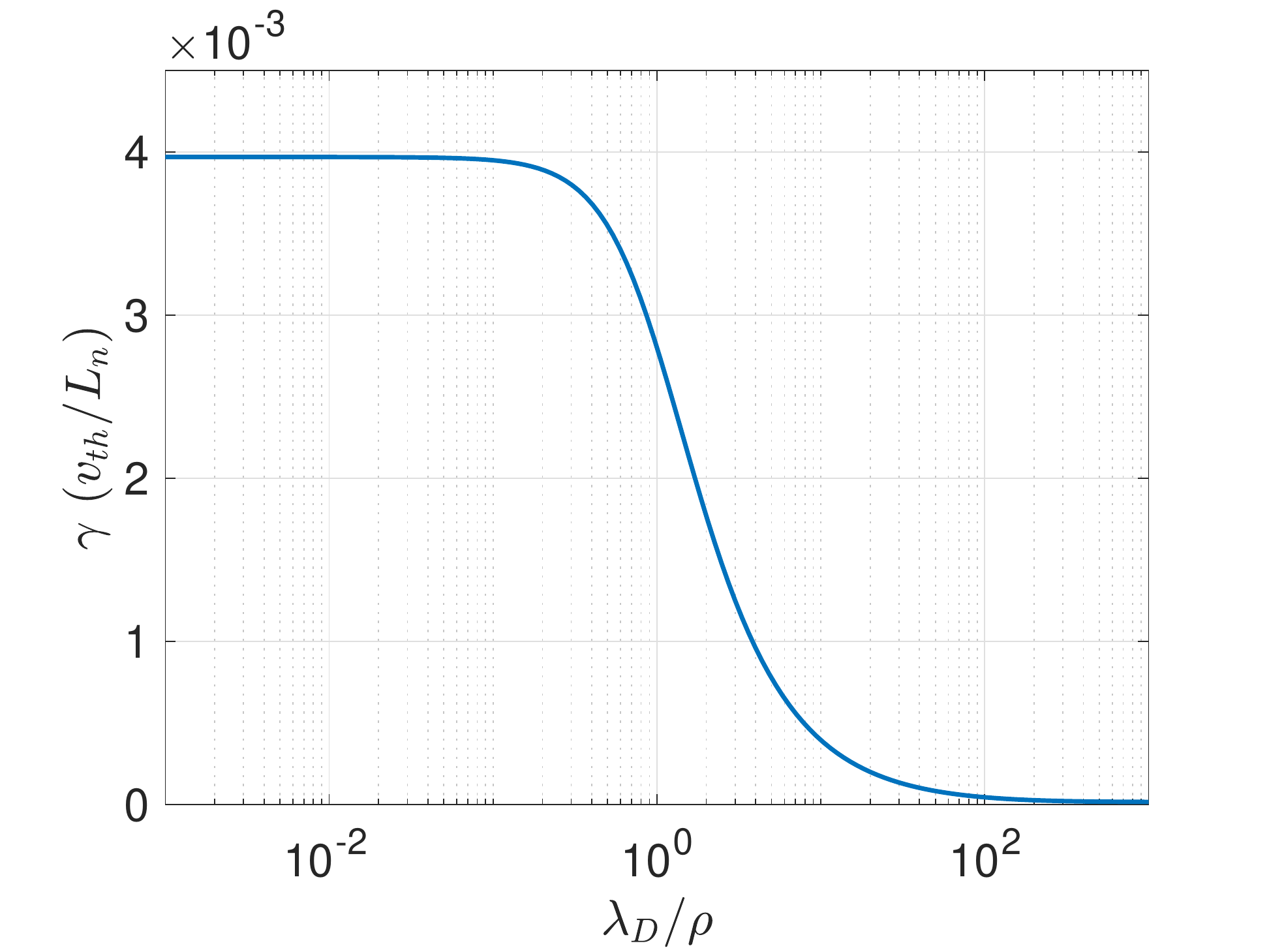}
	\caption{Linear growth rate of the slab universal instability in a pair plasma for $\tau_e=1,\beta=k_\perp \rho=0.1, \eta_p=\eta_e=-0.75$ varies with Debye length $\lambda_D$.}
	\label{fig:ep_lambda}
\end{figure}

The authors would like to remark on the stabilizing effect of Debye shielding in the pair plasma as well. So far our analysis has assumed that the pair plasma is dense enough so that $\lambda_D/\rho_p\ll 1$; however, this condition is usually not satisfied for present-day laboratory produced pair plasmas (e.g., $\rho_p\sim 100~\mu$m while $\lambda_D\sim 3.6$ m in APEX~\cite{hergenhahn2018progress}, and $\rho_p\sim 0.04$ cm while $\lambda_D\sim 1$ cm in the LLNL experiment~\cite{jens2019}) for which the density is low. In our studies Debye shielding effect are always stabilizing, as illustrated in figure~\ref{fig:ep_lambda}. In the case of $\tau_e=1$ (so $\rho_p=\rho_e=\rho$) the Debye shielding effect starts to play a role when $\lambda_D/\rho>0.1$, and has a profound impact on the linear growth rate when $\lambda_D/\rho\sim 1$. This slab $\delta B_\parallel$-universal mode is (almost) completely suppressed when $\lambda_D/\rho\sim 100$. Hence, observation of slab universal instability requires that $\lambda_D/\rho<1$, or equivalently, the plasma number density exceeds a critical value $n_c=B^2/(8\pi m_e c^2)$; $n_c$ is also termed the Brillouin limit.~\cite{brillouin1945theorem} Because Debye shielding stabilizes nearly all instabilities, similar conclusions were drawn in previous studies on pair plasmas, e.g., gyrokinetic GS2 simulations of interchange instability in electron-positron plasma confined in a tokamak configuration~\cite{pedersen2003prospects} and in a linear electrostatic gyrokinetic stability analysis.~\cite{helander2014microstability}

\section{Conclusion}\label{sec:conc}
In this paper, the generalized electromagnetic $\delta B_\parallel$-universal instability dispersion relation in a collisionless slab plasma is derived by performing a gyrokineitic linear stability analysis accounting for electron finite Larmor radius (FLR) and Debye shielding effects. This ion-scale microinstability, driven by density and temperature gradients, vanishes in the long wavelength limit ($k_\perp\to 0$) and typically has a peak growth rate at $k_\perp\rho_i\sim 1$. It also tends to reverse its propagation direction from the electron to the ion diamagnetic direction as $k$ increases. 

We find that the electron FLR effects, at low $k$, generally stabilize the slab $\delta B_\parallel$-universal instability, resulting in a more constrained unstable region in the $\eta_i-\eta_e$ parameter space and reduced linear growth rates. Although this stabilization effect is weak for electron-ion plasmas where the electron gyro-radius is normally two orders of magnitude smaller than ion gyro-radius ($\delta=|\rho_e/\rho_i|=Z/\sqrt{m_eT_e/(m_iT_i)}\sim O(10^{-2})$), it starts to play a role once $\delta>0.1$ (e.g., in cold ion plasma or artificially reducing the ion-electron mass ratio). 
However, electron FLR effects cannot completely stabilize this instability. As a result, we find that this instability persists in the electron-positron pair plasma where the conventional electrostatic universal instability (i.e., drift-wave) is absent due to the unique mass symmetry.
Several interesting features of this instability are observed in the pair plasma. When electrons and positrons are at thermal equilibrium (i.e., $T_p=T_i,L_{T_p}=L_{T_e}$ so $\delta=1$), the slab $\delta B_\parallel$-universal mode becomes a purely growing instability (i.e., $\omega=0$; or the mode stops propagating). As the electron FLR effects exceed the positron (ion) FLR effects ($\delta>1$), it starts to drive, or destabilize, the slab $\delta B_\parallel$-universal instability and to reverse mode propagation direction.
We have also shown that Debye shielding has a strong stabilization effect once the Debye length $\lambda_D$ approaches $\rho_i$, as the electrostatic potential fluctuations $\tilde{\phi}$ associated with the instability are effectively screened.

It is worth pointing out that the gyrokinetic linear analysis of slab universal instability can be further extended, either to the relativistic limit, and to dipole and mirror fields such as those planned for use in confining laboratory-produced pair plasmas.
The nonlinear dynamics of the slab universal instability, and the turbulent transport it induces, remain unexplored research areas. The results may differ from the linear prediction presented here due to the nonlinear excitation of zonal flows and/or electromagnetic fluctuations. We look forward to carrying out local gyrokinetic simulations that shed light on these topics.

\section*{Supplementary Material}
See supplementary material for the complete derivation of the generalized dispersion relation of the slab $\delta B_\parallel$-universal instability.  

\section*{Data Availability Statement}
The data shown in this article are available from the corresponding author upon reasonable request.

\begin{acknowledgments}
B. Zhu and M. Francisquez thank M. J. Pueschel, P. Terry and A. Friedman for the useful discussions and suggestions.
B. Zhu and X. Xu are supported by DOE contract DE-AC52-07NA27344 through the Lawrence Livermore National Laboratory. M. Francisquez is supported by DOE contract DE-FC02-08ER54966. B. Rogers is supported by DOE-SC-0010508. This research used resources of the Discovery cluster supported by the Research Computing Group at Dartmouth College.
\end{acknowledgments}

\appendix

\section{Low-$\beta$ long wavelength limit}\label{app:lowbeta}
%In the low-$\beta$ limit, generalized dispersion relation is analytically attainable.

%In the long wavelength (low $k$) limit, dispersion relation (\ref{eq:dr0}) is analytically obtainable. Here are some useful formulas during the derivation, including:
%\begin{equation}
%J_n(-z)=(-1)^n J_n(z),
%\end{equation}
%small argument expansion for $J_{0,1}$
%\begin{equation}
%J_0(z)=1-\frac{z^2}{4}+O(z^4)\ , \quad J_1(z)=\frac{z}{2}-\frac{z^3}{16}+O(z^5),
%\end{equation}
%and general Gaussian integral
%\begin{equation}
%\int _{0}^{\infty}dx \ e^{-x^2}\left(x+a x^3+bx^5+cx^7+dx^9\right)=\frac{1}{2}+\frac{1}{2}a+ %b+ 3c +12d\ .
%\end{equation}

In the low $\beta$, long wavelength limit, let $u{=}2\omega/k$, then to $O(\beta^2)$
\begin{eqnarray}
\frac{1}{\omega_{bi}}&=&\frac{1}{\omega}\left[1-\frac{\beta_i\alpha_0}{2u}v^2+\frac{\beta_i^2\alpha_0^2}{4u^2}v^4+O(\beta_i^3)\right]\ , \\
\frac{1}{\omega_{be}}&=&\frac{1}{\omega}\left[1+\frac{\beta_e\alpha_0}{2u}v^2+\frac{\beta_e^2\alpha_0^2}{4u^2}v^4+O(\beta_e^3)\right] \,
\end{eqnarray}
so that
\begin{equation}
\begin{split}
\frac{\bar{\omega}_i}{\omega_{bi}}&=\left( 1-\frac{1}{u}+\frac{\eta_i}{u}\right)-\left[\frac{\beta_i\alpha_0}{2u}\left( 1-\frac{1}{u}+\frac{\eta_i}{u}\right)+\frac{\eta_i}{u}\right]v^2\\
&+\left[\frac{\beta_i^2\alpha_0^2}{4u^2}\left( 1-\frac{1}{u}+\frac{\eta_i}{u}\right)+\frac{\beta_i\alpha_0\eta_i}{2u^2}\right]v^4-\frac{\beta_i^2\alpha_0^2\eta_i}{4u^3}v^6+...\ ,
\end{split}
\end{equation}
\begin{equation}
\begin{split}
\frac{\bar{\omega}_e}{\omega_{be}}&=\left( 1+\frac{Z\tau_e}{u}-\frac{Z\tau_e\eta_e}{u}\right)+\left[\frac{\beta_e\alpha_0}{2u}\left( 1+\frac{Z\tau_e}{u}-\frac{Z\tau_e\eta_e}{u}\right)+\frac{Z\tau_e\eta_e}{u}\right]v^2 \\ &+\left[\frac{\beta_e^2\alpha_0^2}{4u^2}\left( 1+\frac{Z\tau_e}{u}-\frac{Z\tau_e\eta_e}{u}\right)+\frac{Z\beta_e\alpha_0\tau_e\eta_e}{2u^2}\right]v^4+\frac{Z\beta_e^2\alpha_0^2\tau_e\eta_e}{4u^3}v^6+...\ .
\end{split}
\end{equation}
Define
\begin{equation}
\alpha_{1i,e}=1+\eta_{i,e}\ , \quad \alpha_{2i,e}=1+2\eta_{i,e}\ , \quad  \alpha_{3i,e}=1+3\eta_{i,e}\ , \quad \alpha_4=1-Z^2\tau_e^2+2\left(\eta_i-Z^2\tau_e^2\eta_e\right),
\end{equation}
then to $O(\beta)$ and $O(k^2)$ order,
\begin{eqnarray}
I_{\phi Q}&=&-\frac{Zk^2}{2}\left[ 1-\frac{\alpha_{1i}}{u}+\frac{1}{\mu} \left(1+\frac{Z\tau_e\alpha_{1e}}{u} \right) \right] -k^2\frac{\lambda_{De}^2}{\rho_i^2}\tau_i \nonumber \\
&+& \frac{Z\beta_i\alpha_0}{2u}\left\{\frac{\alpha_0}{u}+k^2\left[ 1-\frac{\alpha_{2i}}{u}-\frac{Z\tau_e}{\mu}\left( 1+\frac{Z\tau_e\alpha_{2e}}{u} \right) \right] \right\} \nonumber \\
&+& O(\beta^2)+O(k^4)\ , \\
I_{BA}&=&1+\beta_i\left(1+Z\tau_e-\frac{\alpha_4}{u}\right)-\frac{3}{4}\beta_i k^2 \left[ 1-\frac{\alpha_{3i}}{u} +\frac{Z^3\tau_e^2}{\mu}\left( 1+\frac{Z\tau_e\alpha_{3e}}{u}\right) \right] \nonumber \\
&+&O(\beta^2)+O(k^4)\ , \\
I_{\phi A}&=& Z\beta_i \left\{ \frac{\alpha_0}{2u}+
\frac{3}{8}k^2\left[ \left( 1-\frac{\alpha_{2i}}{u}\right)-\frac{Z\tau_e}{\mu}\left(1+\frac{Z\tau_e\alpha_{2e}}{u} \right) \right] \right\} \nonumber \\
&+& O(\beta^2)+O(k^4)\ .
\end{eqnarray}
Substituting above equations to Equation~(\ref{eq:dr0}), defining $\bar{\lambda}_{D_e}^2=2\lambda_{D_e}^2\tau_i/(Z\rho_i^2)$, assuming $\omega{\propto}k$ and only keeping to $O(\beta)$, $O(Z)$ and $O(k^2)$ order yields
\begin{equation} \label{eq:lowkq}
a_2u^2+a_1u+a_0=0
\end{equation}
with
\begin{eqnarray}
a_2&=&\left(1+\frac{1}{\mu}+\bar{\lambda}_{De}^2\right)\left(1+\beta\right)=\left(1+\frac{1}{\mu}+\bar{\lambda}_{De}^2\right)\left[1+\beta_i\left(1+Z\tau_e \right) \right]\ , \\
a_1&=&-\alpha_{1i}\left(1+\beta\right)+\beta_i\left( \frac{\alpha_0}{2}-\alpha_4-\bar{\lambda}_{De}^2\alpha_4 \right) +\frac{1}{\mu}\left[ Z\tau_e\alpha_{1e}\left(1+\beta\right)-\beta_i\left( \frac{Z\tau_e\alpha_0}{2}+\alpha_4 \right)\right]\ , \\
a_0&=&\beta_i\left[ \alpha_{1i}\alpha_4-\frac{\alpha_0\alpha_{2i}}{2}-\frac{Z\tau_e}{\mu}\left( \alpha_{1e}\alpha_4+\frac{Z\tau_e\alpha_0\alpha_{2e}}{2}\right) \right]\ .
\end{eqnarray}

%[Check 2]
%With $Z=1$ and $\mu\to \infty$, above coefficients reproduce Equation (33) of Reference [1]. \\

The instability condition of entropy modes under this limit thus becomes
\begin{equation}
\Delta=a_1^2-4a_0a_2<0
\end{equation}

%\nocite{*}
\bibliography{zhu20ep}% Produces the bibliography via BibTeX.

\newpage
\begin{center}
{\large Supplement material for $\delta B_\parallel$-universal instability analysis} \\
\vspace{0.2cm}
{\small Ben Zhu$^1$, Manaure Francisquez$^2$, Barrett N. Rogers$^3$ and Xue-qiao Xu$^1$} \\
\vspace{0.2cm}
{\footnotesize $^1$LLNL, $^2$MIT, $^3$Dartmouth College}
\end{center}

%\section{Derivation of generalized dispersion relation} \label{app:gdr}
The derivation of generalized dispersion relation of slab universal instability follows similar procedure described in reference~\cite{rogers2018gyrokinetic} (which followed the linear gyrokinetic stability analysis proposed in reference~\cite{antonsen1980kinetic}). In fact, the analysis on the Valsov equation in appendix of reference~\cite{rogers2018gyrokinetic} is valid in our derivation up to Equation~(\ref{eq:geq2}) (Equation (A.32) in reference~\cite{rogers2018gyrokinetic}). Therefore, we only briefly outline these procedures prior Equation~(\ref{eq:geq2}) for completeness in this appendix.

Considering a fully ionized quasi-neutral plasma consists of electrons and ions with charge $q_i{=}Ze$ and mass ratio $\mu{=}m_i/m_e$, the total plasma kinetic pressure is $p_0{=}p_{0i}{+}p_{0e}{=}n_{0i}T_{0i}(1{+}Z\tau_e)$ where the quasi-neutral condition $n_{0e}=Zn_{0i}$ is applied and $\tau_e{=}T_{e0}/T_{i0}$. The quasi-neutral condition also ensures $L_{n_i}{=}L_{n_e}{=}L_n$ where the characteristic length of quantity $f$ is defined as $L_f=f/f'$ (here prime denotes derivative along $x$ direction). Hence, the total balance condition $p_0{+}B_0^2/(8\pi){=}\text{constant}$ implies that $L_B^{-1}{=}-\beta L_p^{-1}/2$, or
\begin{equation}
\frac{L_{n_i}}{L_B}=-\frac{\beta_i\alpha_0}{2}
\end{equation}
with
\begin{equation}
\alpha_0=1+\eta_i+Z\tau_e+Z\tau_e\eta_e\ .
\end{equation}

Further assuming the guiding magnetic field aligns in $z$ direction $\boldsymbol{B}=(B_0(x)+\tilde{B})\hat{z}$ where the perturbed magnetic field $\tilde{B}\propto e^{-i\omega t+iky}$, this results an equilibrium (diamagnetic) current $\boldsymbol{J}_0=-cB_0'/(4\pi)\hat{y}$ carried by diamagnetic drifts $v_{d\alpha}=c p_{\alpha 0}'/(n_0 q_\alpha B_0)$. Here and subscript $\alpha$ represents different plasma species.

After defining
\begin{equation}\label{eq:norms1}
v_{t\alpha}^2=\frac{2T_\alpha}{m_\alpha}\ , \quad \Omega_\alpha=\frac{q_\alpha B_0}{m_\alpha c}\ , \quad \bar{v}=\frac{v}{v_{t\alpha}} \ , \quad \bar{v}_{d\alpha}=\frac{v_{d\alpha}}{v_{t\alpha}}\ ,
\end{equation}
we proceed with a solution to the linearized Vlasov equation as an expansion in $\epsilon\sim \rho_\alpha/L$, where $\rho_\alpha=v_{t\alpha}/\Omega_{\alpha}$ and $L$ is a typical equilibrium scale length such as $L_n$, $L_p$, and the resulting equation when the equilibrium electric field is omitted reads
\begin{equation}\label{eq:vlasov1}
\left\{\partial_t+\v{v}\cdot\nabla+\left[\frac{q}{m}{\tilde{\v{E}}}
+\Omega\left(1+\frac{\tilde{B}}{B_0}\right)
\v{v}\times\v{\hat{z}}\right]\cdot \nabla_v\right\}
\left(F_{eq}+ \tilde{f}\right)=0
\end{equation}
where the equilibrium distribution function which is also expanded in $\epsilon$: $F_{eq}=F_0+F_1+\dots$ obeys
\begin{equation}\label{eq:vlasoveq}
\left(\v{v}\cdot\nabla +\Omega\v{v}\times\v{\hat{z}}\cdot \nabla_v\right)\left(F_0+F_1+\dots\right)=0 \ .
\end{equation}
To zeroth and first order in $\epsilon$, Equation (\ref{eq:vlasoveq}) yields
\begin{equation}
    F_0=\left(\frac{n_0}{\pi^{3/2}v_t^3}\right)e^{-\bar{v}^2}\ ,\quad
    F_1=\frac{v_y}{\Omega L_F}F_0\,
\end{equation}
where $L_F^{-1}=L_n^{-1}+L_T^{-1}(\bar{v}^2-3/2)$.
The perturbation $\tilde{f}\propto e^{-i\omega t+iky}$ obeys
\begin{equation}\label{eq:f}
\left(-i\omega+\v{v}\cdot\nabla+\Omega\v{v}\times\v{\hat{z}}\cdot \nabla_v\right)\tilde{f} +\left(\frac{q}{m}{\tilde{\v{E}}}
+\Omega\frac{\tilde{B}}{B_0}
\v{v}\times\v{\hat{z}}\right)\cdot \nabla_v\left(F_0+ F_1\right)=0.
\end{equation}
Note $\tilde{\v{E}}=-\nabla\phi - \partial_t\v{A}/c$ and define $L_{FT}^{-1}=L_F^{-1}-L_T^{-1}$, the second term in Equation~(\ref{eq:f}) to $O(\epsilon)$
\begin{equation}\label{eq:eeq1}
-i k \frac{q\tilde{\phi}}{m}
\frac{\partial}{\partial {v_y}}
\left(F_0+F_1\right)=i k \tilde{\varphi}\left[
v_y+\frac{v_y^2}{\Omega L_{FT}}-\frac{v_t^2}{2\Omega L_F}
\right]\ ,
\end{equation}
\begin{equation}\label{eq:eeq2}
-\frac{\omega\Omega}{k}\frac{\tilde{B}}{B_0}\frac{\partial}{\partial{v_x}}\left(F_0+F_1\right)=
\frac{2\omega \bar{v}_x\Omega}{k v_t}\tilde{b}+O(\epsilon^2)
\end{equation}
with $\tilde{\varphi}=q\tilde{\phi}F_0/T$ and $\tilde{b}=\tilde{B}F_0/B_0$.
Defining
\begin{equation}\label{eq:hats}
\tilde{\varphi}=\hat{\varphi}e^{iky-i\omega t}\ ,\  \ \tilde{b}=\hat{b}e^{iky-i\omega t}\ ,\  \ \tilde{f}=\hat{f}e^{iky-i\omega t}
\end{equation}
and expanding $\hat{f}=\hat{f}_0+\hat{f}_1+\dots$, 
Equation~(\ref{eq:f}) to $O(\epsilon)$ is
\begin{equation}\label{eq:f2}
\left(-\frac{i \omega}{\Omega}+\frac{v_x\partial_x}{\Omega}+\frac{ik v_y}{\Omega}-\frac{\partial}{\partial \xi}\right)\left(\hat{f}_0+ \hat{f}_1\right)
+\frac{ik\hat{\varphi}}{\Omega}\left[
v_y+\frac{v_y^2}{\Omega L_{FT}}-\frac{v_t^2}{2\Omega L_F}\right] +\hat{b}\bar{v}_x\left(\frac{2\omega }{k v_t}-\frac{v_t}{\Omega L_F}\right)=0\ .
\end{equation}
To leading order
\begin{equation}\label{eq:lead}
\left(\frac{ik v_y}{\Omega}-\frac{\partial}{\partial \xi}\right)\hat{f}_0+\frac{ikv_y}{\Omega}\hat{\varphi}=0\ ,
\end{equation}
thus,
\begin{eqnarray}
\hat{f}_0=-\hat{\varphi}+g e^{-ikv_x/\Omega}
\end{eqnarray}
where $g$ is independent of gyro-angle $\xi$ and is to be determined. At first order
\begin{equation}\label{eq:first}
\left(-\frac{i \omega}{\Omega}+\frac{v_x\partial_x}{\Omega}\right)\hat{f}_0
+\left(\frac{ik v_y}{\Omega}-\frac{\partial}{\partial \xi}\right)\hat{f}_1 
+\frac{ik\hat{\varphi}}{\Omega}\left[
\frac{v_y^2}{\Omega L_{FT}}-\frac{v_t^2}{2\Omega L_F}\right] 
+\hat{b}\bar{v}_x\left(\frac{2\omega}{k v_t}-\frac{v_t}{\Omega L_F}\right)=0\ .
\end{equation}
Multiplying Equation~(\ref{eq:first}) by $e^{ikv_x/\Omega}$ and taking the gyro-angle average, it becomes
\begin{equation}\label{eq:gyro1}
\begin{split}
\int_{-\pi}^\pi &\frac{d\xi}{2\pi}e^{ikv_x/\Omega}\Bigg\{
-\frac{i \omega}{\Omega}\left(-\hat{\varphi}+g e^{-ikv_x/\Omega}\right)
+\frac{v_x}{\Omega}\partial_x\left(-\hat{\varphi}+g e^{-ikv_x/\Omega}\right) +
\frac{ik\hat{\varphi}}{\Omega}
\left[\frac{v_y^2}{\Omega L_{FT}}-\frac{v_t^2}{2\Omega L_F}\right] \\
&+\hat{b}\bar{v}_x\left(\frac{2\omega }{k v_t}-\frac{v_t}{\Omega L_F}\right)\Bigg\}=0
\end{split}
\end{equation}
which eventually leads to
\begin{equation}\label{eq:geq2}
g=\frac{\bar{\omega}}{\omega_b}
\left(J_0\hat{\varphi}+\frac{2\Omega \bar{v}_\perp }{k v_t} J_1\hat{b}\right),
\end{equation}
where
\begin{equation}
\bar{\omega}=\omega-\frac{kv_t^2}{2\Omega L_F}\ ,\quad \omega_b=\omega-\frac{kv_\perp^2}{2\Omega L_B}
\end{equation}
and the normalized perturbed distribution (with respect to $F_0$) thus is
\begin{equation}
\bar{f}_0=\frac{\tilde{f}_0}{F_0}=\left(e^{-ikv_x/\Omega}J_0\frac{\bar{\omega}}{\omega_b}-1\right)\frac{q \tilde{\phi}}{T_0}+
e^{-ikv_x/\Omega}J_1\frac{\bar{\omega}}{\omega_b}\frac{2\bar{v}_\perp}{k\rho}\frac{\tilde{B}}{B_0}.
\end{equation}
Or, for ions and electrons (denoted by subscript $i,e$):
\begin{eqnarray}
\bar{f}_{0i}&=&\left(e^{-ikv_x/\Omega_i}J_0\frac{\bar{\omega}_i}{\omega_{bi}}-1\right)\frac{Ze \tilde{\phi}}{T_{i0}}+
e^{-ikv_x/\Omega_i}J_1\frac{\bar{\omega}_i}{\omega_{bi}}\frac{2\bar{v}_\perp}{k\rho_i}\frac{\tilde{B}}{B_0},\\
\bar{f}_{0e}&=&\left(e^{-ikv_x/\Omega_e}J_0\frac{\bar{\omega}_e}{\omega_{be}}-1\right)\frac{-e \tilde{\phi}}{T_{e0}}+
e^{-ikv_x/\Omega_e}J_1\frac{\bar{\omega}_e}{\omega_{be}}\frac{2\bar{v}_\perp}{k\rho_e}\frac{\tilde{B}}{B_0}.
\end{eqnarray}
Note in this derivation no assumption of $k\rho_e\ll 1$ is made and the full FLR effects are retained for electrons.

(a) According to Ampere's law, the perturbed magnetic field follows
\begin{equation}\label{eq:amp1}
\begin{split}
\frac{\tilde{B}}{B_0}=&\frac{(\nabla\times \tilde{\v{B}})_x}{ikB_0} \\
=& -\frac{4\pi i}{ckB_0}\sum_{i,e}\int d^3vqv_x\tilde{f}_0\\
=& -\frac{4\pi i e}{ck B_0\pi^{3/2}}\int d^3\bar{v} e^{-\bar{v}^2}\bar{v}_x\left(Zn_{0i}v_{ti}\bar{f}_{0i}-n_{0e}v_{te}\bar{f}_{0e}\right)\\
=& -\frac{4\pi i e n_{0e}}{ck B_0\pi^{3/2}}\int d^3\bar{v} e^{-\bar{v}^2}\bar{v}_x \Bigg\{ v_{ti}\left[\left(e^{-ikv_x/\Omega_i}J_0\frac{\bar{\omega}_i}{\omega_{bi}}-1\right)\frac{Ze \tilde{\phi}}{T_{i0}}+
e^{-ikv_x/\Omega_i}J_1\frac{\bar{\omega}_i}{\omega_{bi}}\frac{2\bar{v}_\perp}{k\rho_i}\frac{\tilde{B}}{B_0} \right] \\
&-v_{te} \left[ \left(e^{-ikv_x/\Omega_e}J_0\frac{\bar{\omega}_e}{\omega_{be}}-1\right)\frac{-e \tilde{\phi}}{T_{e0}}+
e^{-ikv_x/\Omega_e}J_1\frac{\bar{\omega}_e}{\omega_{be}}\frac{2\bar{v}_\perp}{k\rho_e}\frac{\tilde{B}}{B_0} \right] \Bigg\}.
\end{split}
\end{equation}
Collecting terms this maybe written as
\begin{equation}\label{eq:aamp0}
I_{BA}\frac{\tilde{B}}{B_0}=I_{\phi A}\frac{e\tilde{\phi}}{T_{i0}}
\end{equation}
where
\begin{eqnarray}
I_{BA} &=&1+\frac{4\pi i e n_{0e}}{ck B_0\pi^{3/2}} \int d^3\bar{v} e^{-\bar{v}^2}\bar{v}_x
\left(v_{ti}e^{-ikv_x/\Omega_i}J_1\frac{\bar{\omega}_i}{\omega_{bi}}\frac{2\bar{v}_\perp}{k\rho_i}
-v_{te}e^{-ikv_x/\Omega_e}J_1\frac{\bar{\omega}_e}{\omega_{be}}\frac{2\bar{v}_\perp}{k\rho_e} \right), \label{eq:IBA0} \\
I_{\phi A} &=& \frac{-4\pi i e n_{0e}}{ck B_0\pi^{3/2}} \int d^3\bar{v} e^{-\bar{v}^2}\bar{v}_x
\left[Zv_{ti}\left(e^{-ikv_x/\Omega_i}J_0\frac{\bar{\omega}_i}{\omega_{bi}}-1\right)
+v_{te}\left(e^{-ikv_x/\Omega_e}J_0\frac{\bar{\omega}_e}{\omega_{be}}-1\right)\tau_i\right]\ . \label{eq:IphiA0} 
\end{eqnarray}
Here $\tau_i=T_{i0}/T_{e0}=\tau_e^{-1}$. Defining
\begin{equation}\label{eq:f1def}
f_{1\alpha}=\frac{i\Omega_\alpha}{kv_x}\left(e^{-ikv_x/\Omega_\alpha}-1+\frac{ikv_x}{\Omega_\alpha}\right)\ \to\ 
e^{-ikv_x/\Omega_\alpha}=1-\frac{ikv_x}{\Omega_\alpha}\left(1+f_{1\alpha}\right)
\end{equation}
and collecting real terms above equations may be written as
\begin{eqnarray}
I_{BA} &=&1+ \int \frac{d^3\bar{v}}{\pi^{3/2}} \bar{v}_\perp^2 e^{-\bar{v}^2}\bar{v}_x^2
\left[ \beta_i \left( 1+f_{1i} \right)\frac{\bar{\omega}_i}{\omega_{bi}}\frac{2J_1}{k\rho_i \bar{v}_\perp}
+\beta_e \left( 1+f_{1e} \right)\frac{\bar{\omega}_e}{\omega_{be}}\frac{2J_1}{k\rho_e \bar{v}_\perp} \right], \label{eq:IBA1} \\
I_{\phi A} &=& -Z\beta_i \int \frac{d^3\bar{v}}{\pi^{3/2}} e^{-\bar{v}^2}\bar{v}_x^2
\left[ \left( 1+f_{1i} \right)J_0\frac{\bar{\omega}_i}{\omega_{bi}}
-\left( 1+f_{1e} \right)J_0\frac{\bar{\omega}_e}{\omega_{be}} \right]\ .\label{eq:IphiA1}
\end{eqnarray}
Transforming to plane-polar perpendicular velocity components with 
\begin{equation}\label{eq:polar}
\int \frac{d^3\bar{v}}{\pi^{3/2}}=\frac{1}{\pi^{3/2}} \int_{-\infty}^{\infty}d\bar{v}_z \int_{-\infty}^{\infty}d\bar{v}_y  \int_{-\infty}^{\infty}d\bar{v}_x=
2\int _{-\infty}^{\infty}\frac{d\bar{v}_z}{\sqrt{\pi}}\int_0^{\infty}d\bar{v}_\perp \bar{v}_\perp \int_{-\pi}^{\pi} \frac{d\xi}{2\pi}\ \ ,
\end{equation}
using the identity 
\begin{equation}\label{eq:f2i}
\frac{\bar{v}_\perp}{k\rho}J_1\left(k\rho \bar{v}_\perp\right)=\int_{-\pi}^{\pi} \frac{d\xi}{2\pi}\bar{v}_x^2\left(1+f_{1}\right)\ ,
\end{equation}
carrying out the $v_z$ integrals with $\bar{v}^2=\bar{v}^2_z+ \bar{v}^2_\perp$ and noting the Gaussian integral
\begin{equation}
\int_{-\infty}^{\infty}dx e^{-x^2}=\sqrt{\pi}
\end{equation}
%\begin{equation}\label{eq:itable}
%\begin{split}
%\int _{-\infty}^{\infty}\frac{dx}{\sqrt{\pi}}e^{-x^2}\left(1+a x^2+bx^4+cx^6\right)&=1+\frac{1}{2}a+ \frac{3}{4}b+\frac{15}{8}c\ , \\
%\int _{0}^{\infty}dx \ e^{-x^2}\left(x+a x^3+bx^5+cx^7+dx^9\right)&=\frac{1}{2}+\frac{1}{2}a+ b+ 3c +12d\ ,
%\end{split}
%\end{equation}
Equation~(\ref{eq:IBA1}) and (\ref{eq:IphiA1}) become
\begin{eqnarray} 
I_{BA}&=&1+4\int_0^\infty d\bar{v}_{\perp} \bar{v}_{\perp}^3 e^{-\bar{v}_{\perp}^2}\left( J_1^2\frac{\beta_i \bar{\omega}_i}{k^2\rho_i^2\omega_{bi}}+J_1^2\frac{\beta_e \bar{\omega}_e}{k^2\rho_e^2\omega_{be}}\right)\ ,\label{eq:IBA2}  \\
I_{\phi A}&=&-2Z\beta_i \int_0^\infty d\bar{v}_{\perp} \bar{v}_{\perp}^2 e^{-\bar{v}_{\perp}^2}\left( J_0J_1\frac{ \bar{\omega}_i}{k\rho_i \omega_{bi}} -J_0J_1\frac{ \bar{\omega}_e}{k\rho_e \omega_{be}}\right)\ . \label{eq:IphiA2}
\end{eqnarray}

(b) The quasi-neutrality condition ($Zn_i{=}n_e$) is now replaced by Gauss's law in order to include Debye shielding effect. To the required order it is written as
\begin{equation}\label{eq:qn1}
\begin{split}
\nabla^2\tilde{\phi}&=-4\pi e\int d^3{v}\left(Z\tilde{f}_{0i}-\tilde{f}_{0e}\right) \\
&=-4\pi e\int \frac{d^3\bar{v}}{\pi^{3/2}}e^{-\bar{v}^2}\left(Zn_{i0}\bar{f}_{0i}-n_{0e}\bar{f}_{0e}\right) \\
&=-4\pi en_{0e}\int \frac{d^3\bar{v}}{\pi^{3/2}}e^{-\bar{v}^2}\left(\bar{f}_{0i}-\bar{f}_{0e}\right)\ .
\end{split}
\end{equation}
Therefore,
\begin{equation}\label{eq:qn2}
\begin{split}
k^2\lambda_{De}^2\tau_i\frac{e\tilde{\phi}}{T_{i0}}=\int \frac{d^3\bar{v}}{\pi^{3/2}}e^{-\bar{v}^2}\Bigg\{ \left[\left(e^{-ikv_x/\Omega_i}J_0\frac{\bar{\omega}_i}{\omega_{bi}}-1\right)\frac{Ze \tilde{\phi}}{T_{i0}}+
e^{-ikv_x/\Omega_i}J_1\frac{\bar{\omega}_i}{\omega_{bi}}\frac{2\bar{v}_\perp}{k\rho_i}\frac{\tilde{B}}{B_0} \right]\\
-\left[\left(e^{-ikv_x/\Omega_e}J_0\frac{\bar{\omega}_e}{\omega_{be}}-1\right)\frac{-e \tilde{\phi}}{T_{e0}}+
e^{-ikv_x/\Omega_e}J_1\frac{\bar{\omega}_e}{\omega_{be}}\frac{2\bar{v}_\perp}{k\rho_e}\frac{\tilde{B}}{B_0} \right]
\Bigg\}
\end{split}
\end{equation}
where electron Debye length $\lambda_{De}=\sqrt{T_e/(4\pi e^2n_{0e})}$. Collecting terms this becomes
\begin{equation}\label{eq:qn3}
I_{\phi Q}\frac{e\tilde{\phi}}{T_{i0}}=I_{BQ}\frac{\tilde{B}}{B_0}
\end{equation}
with
\begin{eqnarray}
I_{\phi Q}&=&\int \frac{d^3\bar{v}}{\pi^{3/2}}e^{-\bar{v}^2}\left[
Z \left(e^{-ikv_x/\Omega_i}J_0\frac{\bar{\omega}_i}{\omega_{bi}}-1 \right)
+\left(e^{-ikv_x/\Omega_e}J_0\frac{\bar{\omega}_e}{\omega_{be}}-1\right)\tau_i\right]-k^2\lambda_{De}^2\tau_i\ , \label{eq:iphiq0}\\
I_{B Q}&=&-\int \frac{d^3\bar{v}}{\pi^{3/2}}e^{-\bar{v}^2}\bar{v}_\perp^2\left(
e^{-ikv_x/\Omega_i}\frac{2J_1}{k\rho_i\bar{v}_\perp}\frac{\bar{\omega}_i}{\omega_{bi}}
-e^{-ikv_x/\Omega_e}\frac{2J_1}{k\rho_e\bar{v}_\perp}\frac{\bar{\omega}_e}{\omega_{be}}\right)\ . \label{eq:bq0}
\end{eqnarray}
Following the same steps as before, and noting that
\begin{equation}
J_0(k\rho\bar{v}_\perp)=\int_{-\pi}^{\pi}\frac{d\xi}{2\pi}e^{\pm ikv_x/\Omega}
\end{equation}
one obtains
\begin{eqnarray}
I_{\phi Q}&=&2\int_0^\infty d\bar{v}_{\perp} \bar{v}_{\perp} e^{-\bar{v}_{\perp}^2}\left[
Z\left(J_0^2\frac{\bar{\omega}_i}{\omega_{bi}}-1\right)+\left(J_0^2\frac{\bar{\omega}_e}{\omega_{be}}-1\right)\tau_i
\right]-k^2\lambda_{De}^2\tau_i\ ,\\
I_{BQ}&=&-4 \int_0^\infty d\bar{v}_{\perp} \bar{v}_{\perp}^2 e^{-\bar{v}_{\perp}^2}\left( J_0J_1\frac{ \bar{\omega}_i}{k\rho_i \omega_{bi}} -J_0J_1\frac{ \bar{\omega}_e}{k\rho_e \omega_{be}}\right)
\end{eqnarray}
with the aid of Equation~(\ref{eq:IphiA2}), implying with Equation~(\ref{eq:aamp0}) the dispersion relation:
\begin{equation}\label{eq:dr0}
I_{\phi Q} I_{BA}=I_{BQ}I_{\phi A}=2\left(I_{\phi A}\right)^2/(Z\beta_i)\ .
\end{equation}
%Note that the $L_F$ expression (\ref{eq:F1}) embedded in $\bar{\omega}$ is also modified after the $v_z$ integration:
%\begin{equation}
%\frac{1}{L_F}=\frac{1}{L_n}+\frac{1}{L_T}\left(\bar{v}_\perp^2-1\right).
%\end{equation}

(c) Normalizing frequencies to $v_{ti}/L_n$, lengths to $\rho_i$ (so for example $\omega=\omega_{phys} L_n/v_{ti}$ and $k=k_{phys}\rho_i$) and denoting $\bar{v}_\perp$ as $v$ for simplicity, the expressions for $I_{\phi Q}$, $I_{BA}$ and $I_{\phi A}$ become
\begin{eqnarray}
I_{\phi Q}&=&2\int_0^\infty dv v e^{-v^2}\left[
Z\left(J_0^2\frac{\bar{\omega}_i}{\omega_{bi}}-1\right)+\left(J_0^2\frac{\bar{\omega}_e}{\omega_{be}}-1\right)\tau_i\right] - k^2\frac{\lambda_{De}^2}{\rho_i^2}\tau_i \ , \label{eq:IphiQ}\\
I_{BA}&=&1+4\int_0^\infty dv v^3 e^{-v^2}\left( J_1^2\frac{\beta_i}{k^2}\frac{\bar{\omega}_i}{\omega_{bi}}+\frac{\mu}{Z^2\tau_e}J_1^2\frac{\beta_e}{k^2}\frac{\bar{\omega}_e}{\omega_{be}}\right)\ ,\label{eq:IBA}  \\
I_{\phi A}&=&-2Z\beta_i \int_0^\infty dv v^2 e^{-v^2}\left( J_0J_1\frac{1}{k}\frac{ \bar{\omega}_i}{\omega_{bi}} +\frac{\mu^{1/2}}{Z\tau_e^{1/2}}J_0J_1\frac{1}{k}\frac{ \bar{\omega}_e}{\omega_{be}}\right)\ \label{eq:IphiA}
\end{eqnarray}
with
\begin{eqnarray}
\bar{\omega}_i=\omega-\frac{k}{2}\left[ 1+\eta_i \left(v^2-1\right)\right] \ , \quad \omega_{bi}= \omega+\frac{\beta_i\alpha_0}{4}kv^2 , \\
\bar{\omega}_e=\omega+\frac{Z\tau_e k}{2}\left[ 1+\eta_e \left(v^2-1\right)\right] \ , \quad \omega_{be}=\omega-\frac{\beta_e\alpha_0}{4}kv^2
\end{eqnarray}
and the arguments of Bessel function $J_{0,1}$ for ions and electrons are $kv$ and $-Z\tau_e^{1/2}kv/\mu^{1/2}$ respectively.

\end{document}